\documentclass[longauth]{aa}

\usepackage{graphicx}
\usepackage{txfonts}
\usepackage[colorlinks=true,citecolor=blue,linkcolor=blue]{hyperref}
\usepackage{natbib}
\bibpunct{(}{)}{;}{a}{}{,} % to follow the A&A style
\usepackage{bm}
\usepackage{upgreek}
\usepackage[dvipsnames]{xcolor}
\usepackage{amsmath}    % Advanced maths commands
\newcommand{\ho}{H_0}
\newcommand{\Ho}{$H_0$}
\newcommand{\lcdm}{$\Lambda$CDM}
\newcommand{\lensname}{RXJ1131$-$1231}

\newcommand{\hunit}{km s$^{-1}$ Mpc$^{-1}$}
\newcommand{\kmps}{km s$^{-1}$}
\newcommand{\Dd}{D_{\rm d}}
\newcommand{\Ds}{D_{\rm s}}
\newcommand{\Dds}{D_{\rm ds}}
\newcommand{\Ddt}{D_{\Delta t}}
\newcommand{\balpha}{\bm{\alpha}}

\newcommand{\bvarsigma}{\bm{\varsigma}}

\newcommand{\btheta}{\bm{\theta}}
\newcommand{\rmd}{{\rm d}}
\newcommand{\moment}[1]{\langle{#1}\rangle}
\newcommand{\pml}{$\pm$}

\newcommand{\secref}[1]{Section~\ref{#1}}
\newcommand{\figref}[1]{Figure~\ref{#1}}
\newcommand{\tabref}[1]{Table~\ref{#1}}
\newcommand{\equref}[1]{Equation~(\ref{#1})}

\newcommand{\hval}{77.1_{-7.1}^{+7.3}}
\newcommand{\Ddval}{865_{-81}^{+85}}
\newcommand{\Ddtval}{2180_{-271}^{+472}}
\newcommand{\deltabicani}{3.5}
\newcommand{\deltabicpl}{3.8}
\newcommand{\chenhval}{$78.3^{+3.4}_{-3.3}$}
\newcommand{\birrerhval}{$74.5^{+5.6}_{-6.1}$}
\newcommand{\birrerhvalalt}{$73.3^{+5.8}_{-5.8}$}
\newcommand{\suyuhval}{$80.0_{-4.7}^{+4.5}$}
\newcommand{\birrerhvalthislens}{$74.5_{-7.8}^{+8.0}$}
\newcommand{\birrerhvalthislensalt}{$86.6_{-6.9}^{+6.8}$}

\newcommand{\millonhval}{$74.2_{-1.6}^{+1.6}$}
\newcommand{\ajsfou}[1]{{{#1}}}
\newcommand{\ajstwo}[1]{{{#1}}}
\newcommand{\ajsthr}[1]{{{#1}}}
\newcommand{\ajsfiv}[1]{{{#1}}}

\begin{document} 

    \title{TDCOSMO}
    
    \subtitle{XII. Improved Hubble constant measurement from lensing time delays using spatially resolved stellar kinematics of the lens galaxy\thanks{Reduced Keck Cosmic Web Imager data analyzed in this paper are also available at the CDS via anonymous ftp to \href{https://cdsarc.cds.unistra.fr/}{cdsarc.cds.unistra.fr} (ftp://130.79.128.5) or via \url{https://cdsarc.cds.unistra.fr/viz-bin/cat/J/A+A/673/A9}. \textsc{jupyter} notebooks and \textsc{python} scripts used in this analysis are available at \url{https://github.com/TDCOSMO/RXJ1131_KCWI/}.} 
    }
    
    \titlerunning{\Ho\ from spatially resolved kinematics of \lensname}
    \authorrunning{A.~J.~Shajib et al.}
    
    %\subtitle{Breaking the MST in TDC in generic cosmologies}
    \author{
    	Anowar~J.~Shajib,\inst{1, 2}\fnmsep\thanks{NFHP Einstein Fellow}
    	Pritom~Mozumdar,\inst{3}
	    Geoff~C.-F.~Chen,\inst{4} %\fnmsep\thanks{Current email address: gcfchen@astro.ucla.edu}
    	Tommaso~Treu,\inst{4}
   		Michele~Cappellari,\inst{5}
	    Shawn Knabel,\inst{4}
    	Sherry~H.~Suyu,\inst{6, 7, 8}
    	Vardha~N.~Bennert,\inst{9}
    	Joshua~A.~Frieman,\inst{1, 2, 10}
    	Dominique~Sluse,\inst{11}
    	Simon~Birrer,\inst{12, 13, 14}
    	Frederic~Courbin,\inst{15}
    	Christopher~D.~Fassnacht,\inst{3}
    	Lizvette~Villafa\~na,\inst{4}
    	Peter~R.~Williams\inst{4}
    }
    
    \institute{
    Department  of  Astronomy  \&  Astrophysics,  University  of Chicago, Chicago, IL 60637, USA; \email{ajshajib@uchicago.edu}
    \and
    Kavli Institute for Cosmological Physics, University of Chicago, Chicago, IL 60637, USA
    \and
    Department of Physics and Astronomy, University of California, Davis, CA 95616, USA
    \and
    Department of Physics and Astronomy, University of California, Los Angeles, CA 90095, USA
    \and
    Sub-Department of Astrophysics, Department of Physics, University of Oxford, Denys Wilkinson Building, Keble Road, Oxford, OX1 3RH, UK
    \and
    Technical University of Munich, TUM School of Natural Sciences, Department of Physics, James-Franck-Str.~1, Garching, 85748, Germany
    \and
    Max Planck Institute for Astrophysics, Karl-Schwarzschild-Str.~1, Garching, 85748, Germany
    \and
    Institute of Astronomy and Astrophysics, Academia Sinica, 11F of ASMAB, No.1, Section 4, Roosevelt Road, Taipei, 10617, Taiwan
    \and
    Physics Department, California Polytechnic State University, San Luis Obispo, CA 93407, USA
    \and
    Fermi National Accelerator Laboratory, P.O.\ Box 500, Batavia, IL 60510, USA
    \and
    STAR Institute, Quartier Agora, All\'ee du Six Ao\^ut, 19c, 4000 Li\'ege, Belgium
    \and
    Kavli Institute for Particle Astrophysics and Cosmology and Department of Physics, Stanford University, Stanford, CA 94305, USA
    \and
    SLAC National Accelerator Laboratory, Menlo Park, CA, 94025
    \and
    Department of Physics and Astronomy, Stony Brook University, Stony Brook, NY 11794, USA
    \and
    Institute of Physics, Laboratory of Astrophysics, Ecole Polytechnique F\'ed\'erale de Lausanne (EPFL), Observatoire de Sauverny, 1290 Versoix, Switzerland
    }
             
   \date{Received xxx, xxxx; accepted xxx, xxxx}

% \abstract{}{}{}{}{} 
% 5 {} token are mandatory
 
  \abstract
  {
     Strong-lensing time delays enable measurement of the Hubble constant ($H_{0}$) independently of other traditional methods. The main limitation to the precision of time-delay cosmography is mass-sheet degeneracy (MSD). Some of the previous TDCOSMO analyses broke the MSD by making assumptions about the mass density profile of the lens galaxy, reaching 2\% $H_0$ precision from seven lenses. However, this approach could potentially bias the $H_0$ measurement or underestimate the errors. In this work, for the first time, we break the MSD using spatially resolved kinematics of the lens galaxy in \lensname\ obtained from the Keck Cosmic Web Imager spectroscopy, in combination with previously published time delay and lens models derived from  \textit{Hubble} Space Telescope imaging. This approach allows us to robustly estimate \Ho, effectively implementing a maximally flexible mass model. Following a blind analysis, we estimate the angular diameter distance to the lens galaxy $\Dd=\Ddval$ Mpc and the time-delay distance $\Ddt=\Ddtval$ Mpc, giving $\ho=\hval$ \hunit\ -- for a flat $\Lambda$ cold dark matter cosmology. The error budget accounts for all uncertainties, including the MSD inherent to the lens mass profile and the line-of-sight effects, and those related to the mass--anisotropy degeneracy and projection effects. Going from single-aperture to spatially resolved stellar kinematics improves the \Ho\ constraint from 13\% to 9\% for this single lens, using maximally flexible models. Our new measurement is in excellent agreement with those obtained in the past by H0LiCOW using simple parametric models for this single system (\Ho\ = \chenhval\ \hunit), and for seven lenses by TDCOSMO with single-aperture kinematics using the same maximally flexible models used by us (\Ho\ = \birrerhvalalt\ \hunit), corroborating the methodology.
}
  

  \maketitle
%
%-------------------------------------------------------------------
\section{Introduction}

The Hubble constant, \Ho, the current value of the Universe's expansion rate, is a crucial cosmological parameter that also sets the extragalactic distance scale.
Recently, tension has emerged between early- and late-Universe estimates of \Ho ~\citep[e.g.,][]{Freedman21, Abdalla22}. The temperature and polarisation fluctuations in the cosmic microwave background (CMB) provide an estimate of the Hubble parameter at the last scattering surface $H(z \approx 1100)$, which can be extrapolated to the current epoch using the $\Lambda$ cold dark matter (\lcdm) cosmology. The CMB measurements from \textit{Planck} give $H_0 = 67.4 \pm 0.5$ \hunit\ \citep{PlanckCollaboration18} and $H_0 = 67.6 \pm 1.1$ \hunit \citep{Aiola20}. In the local Universe, \Ho\ can be estimated using the cosmic distance ladder, which uses luminosity distances of type Ia supernovae (\ajstwo{SNe Ia}) with their absolute brightness calibrated using different classes of stars. The Supernova \Ho\ for the Equation of State of the dark energy (SH0ES) team uses Cepheids and parallax distances for this calibration, and they find $H_0 = 73.04 \pm 1.04$ \hunit\ \citep{Riess22}. This value is in 5$\sigma$ tension with the \textit{Planck} CMB-based measurements. If this difference is not due to systematic errors in either of these measurements \ajstwo{\citep[e.g.,][]{Efstathiou21}}, then this tension could point to new physics beyond the \lcdm\ cosmological model \citep[e.g.,][]{Knox20}. 

To determine whether this ``Hubble tension'' is due to systematics or new physics, multiple independent methods to measure \Ho\ are needed \ajstwo{\citep[e.g.,][]{Verde19, DiValentino21b, Freedman21}}. The Carnegie--Chicago Hubble Project uses the tip of the red giant branch (TRGB) to calibrate the \ajstwo{SNe Ia} absolute brightness and measures $H_0 = 69.6 \pm 1.9$ \hunit \ \citep{Freedman19, Freedman20}. This TRGB-calibrated measurement is statistically consistent with both the SH0ES measurement and the CMB-based measurements. However, several independent local probes strengthen the ``\Ho\ tension'' by measuring values consistent with the SH0ES value. For example, the Megamaser Cosmology Project (MCP) estimates $H_0 = 73.9 \pm 3.0$ \hunit \citep{Pesce20}, the surface brightness fluctuation (SBF) method measures $H_0 = 73.7 \pm 0.7 \pm 2.4$ \hunit\ \citep{Blakeslee21}, and the Tully--Fisher-relation-based method calibrated with Cepheids measures $H_0 = 75.1 \pm 0.2 \pm 0.3$ \hunit\ \citep{Kourkchi20}.

Strong-lensing time delays provide an independent measurement of \Ho\ \citep[\citealt{Refsdal64}; for an up-to-date review, see][for a historical perspective, see \citealt{Treu16b}]{Birrer22b, Treu22}. In strong lensing, a background source appears as multiple images due to the gravitational deflection of photons by a massive foreground galaxy or galaxy cluster. The photons that were emitted at the same time from the background source arrive in different images with a relative time delay. This time delay carries cosmological information through a combination of angular diameter distances involved in the lensing system. This combination is referred to as the ``time-delay distance", which is inversely proportional to \Ho\ \citep{Refsdal64, Schneider92, Suyu10}. The Time-Delay COSMOgraphy (TDCOSMO) collaboration has analyzed seven time-delay lenses to measure $H_0$ with 2\% error, $\ho = 74.2 \pm 1.6$ \hunit\, assuming a power-law or composite \citep[i.e., stars and Navarro--Frenk--White (NFW) halo;][]{Navarro96, Navarro97} mass profile for the lensing galaxies \citep{Millon20}. The TDCOSMO collaboration encompasses \ajstwo{the COSmological MOnitoring of GRAvItational Lenses \citep[COSMOGRAIL;][]{Courbin05, Millon20b}}, the \Ho\ Lenses in COSMOGRAIL's Wellspring \citep[H0LiCOW;][]{Suyu10, Suyu13, Bonvin17, Birrer19b, Rusu20, Wong20}, the Strong-lensing High Angular Resolution Programme \citep[SHARP;][]{Chen19}, and the STRong-lensing Insights into the Dark Energy Survey \citep[STRIDES;][]{Treu18, Shajib20} collaborations.

The simple parametric lens models, e.g., the power law, adopted in the TDCOSMO analyses are ``industry standard'' consistent with non-lensing measurements. The TDCOSMO collaboration has performed various systematic checks on the adopted lens modeling procedure. These checks find potential systematic biases to be lower than the acceptable limit ($\sim$1\%) from the choice of mass model parametrization \citep[i.e., power law or composite,][]{Millon20}, from ignoring dark substructures in the lens galaxy's halo \citep{Gilman20}, from ignoring disky or boxy-ness in the baryonic distribution \citep{VandeVyvere22}, from using different lens modeling software \citep{Shajib22}, and from ignoring potential isodensity twists and ellipticity gradients in the lens galaxy \citep{VandeVyvere22b}. However, a significant source of potential systematics could arise due to the relatively simple parametrization of the lens mass profile \citep{Kochanek20}. The well-known mass-sheet degeneracy (MSD) does not allow one to constrain the mass profile shape of the deflector galaxy from lens imaging observables alone \citep{Falco85, Schneider13, Schneider14}. Non-lensing observables, such as the deflector galaxy's velocity dispersion or the source's unlensed intrinsic brightness, are required to break the mass-sheet degeneracy and simultaneously constrain \Ho\ and the mass profile shape \citep{Treu02b, Shajib18, Yildirim20, Yildirim21, Birrer20, Birrer22}. 

The TDCOSMO collaboration has redesigned the experiment to mitigate this systematic by relaxing the simple parametric assumptions in the mass profile and constraining the profile shape solely from stellar velocity dispersion measurements of the lensing galaxies \citep{Birrer20}. Relaxing the assumption on the mass profile leads to an increase in the \Ho\ uncertainty from 2 to 8\% -- which is dominated by the uncertainty of the measured velocity dispersion -- giving $\ho = 74.5_{-6.1}^{+5.6} $ \hunit. One approach to improving the precision is to incorporate prior information on the mass profile shape from the measured velocity dispersions of a larger sample of external lenses without measured time delays. Assuming that the Sloan Lens ACS (SLACS) survey's lens galaxies are drawn from the same population as the TDCOSMO lens galaxies and using their velocity dispersions to constrain the mass profile shape, the uncertainty on \Ho\ improves to 5\%, giving $\ho = 67.4^{+4.1}_{-3.2}$ \hunit \citep{Birrer20}. Note that this estimate is statistically consistent within $1\sigma$ with the larger 8\% \Ho\ measurement above. However, the shift could also arise from systematic differences, \ajsthr{e.g., a difference between the parent populations of time-delay and non-time-delay lenses \citep{Gomer22}}. Such differences could arise, for example, from evolutionary effects, as the SLACS sample is at lower redshift than the TDCOSMO lenses \citep[see, e.g.,][for a discussion of the evolution of mass density profiles of massive elliptical galaxies]{Sonnenfeld15}.

Spatially resolved velocity dispersion measurements of lens galaxies for systems with measured time delays are critical to drastically improving the \Ho\ precision, given the limited sample size of time-delay lenses \citep{Shajib18, Yildirim21}. The spatially resolved nature of the measured velocity dispersion is especially powerful in simultaneously breaking the MSD and the mass-anisotropy degeneracy \ajstwo{\citep{Cappellari08, Barnabe09, Barnabe12, Collett18, Shajib18}}. Spatially resolved velocity dispersion measurements for $\sim$40 time-delay lens galaxies will yield an independent $\lesssim$2\% \Ho\ measurement without any mass profile assumption \citep{Birrer21c}. Additional constraints from velocity dispersion measurements of non-time-delay lens galaxies or magnification information for standardizable lensed type Ia supernovae can further improve the uncertainty to $\lesssim1$\% \citep{Birrer21c, Birrer22}.

In this paper, we measure the spatially resolved velocity dispersion for the lens galaxy in the strongly lensed quasar system \lensname using the Keck Cosmic Web Imager (KCWI) integral field spectrograph on the W.~M.~Keck Observatory \citep{Morrissey12, Morrissey18}.  and constrain \Ho\ without any mass profile assumption from this single time-delay lens system. This is the first application of spatially resolved velocity dispersion from a time-delay lens to measure \Ho. This lens system was previously used to measure \Ho\ by combining the observed imaging data, single-aperture velocity dispersion, time delays, and analysis of the line-of-sight environment \citep{Suyu13, Suyu14}. However, \ajsfou{these previous studies} assumed simple parametrizations for the mass profile, such as a power law or a combination of the NFW profile and the stellar profile with constant mass-to-light, \ajsfou{which is the industry standard in modeling of galaxy-scale lenses \citep{Shajib22c}}. \ajsthr{\citet{Birrer16} marginalized over the MSD effect for the system \lensname\ to constrain \Ho\ using a single-aperture velocity dispersion measurement.} Here, we allow the \ajsfou{maximal freedom} in the MSD \ajsthr{by introducing one free parameter on top of the simply parametrized mass profile constrained by lens modeling, which is completely degenerate with \Ho.} 

This paper is organized as follows. In \secref{sec:obs_DRP}, we describe the observational strategy and data reduction. In \secref{sec:kinematics}, we describe the procedures to extract the spatially resolved kinematics map from the KCWI data. In \secref{sec:theory}, we briefly review the lensing and dynamical formalisms and how we combine the two to mitigate the MSD in our analysis. Then in \secref{sec:dynamic_models}, we describe our dynamical models and present results. We infer the cosmological parameters from our analysis in the \secref{sec:cosmo_inference}. We discuss our results in \secref{sec:discussion} and conclude the paper in \secref{sec:conclusion}.

\ajsfou{We performed the cosmological inference blindly in this paper. The measurement of velocity dispersion was not blinded. However, we blinded the cosmological and other model parameters directly related to cosmological parameters in the dynamical modeling. \ajsfiv{Before unblinding, this analysis went through an internal collaboration-wide review and code review. After all the coauthors had agreed that the necessary systematic checks were satisfactorily performed,} we froze the analysis and unblinded on 5 January 2023. All the sections in this paper except for the final discussion in \secref{sec:discussion} and summary in \secref{sec:conclusion} were written before unblinding. After unblinding, we only made minor edits for clarity and grammatical corrections in the previous sections and added the unblinded numbers where relevant in the abstract, main text, and plots.}

\section{Observations and data reduction}
\label{sec:obs_DRP}

In this section, we provide a brief description of the lens system \lensname\ (\secref{sec:lens_description}), the spectroscopic observation with KCWI (\secref{sec:kcwi_spectra}), and the data reduction procedure (\secref{sec:reduction}).

\begin{figure}
	\includegraphics[width=\columnwidth]{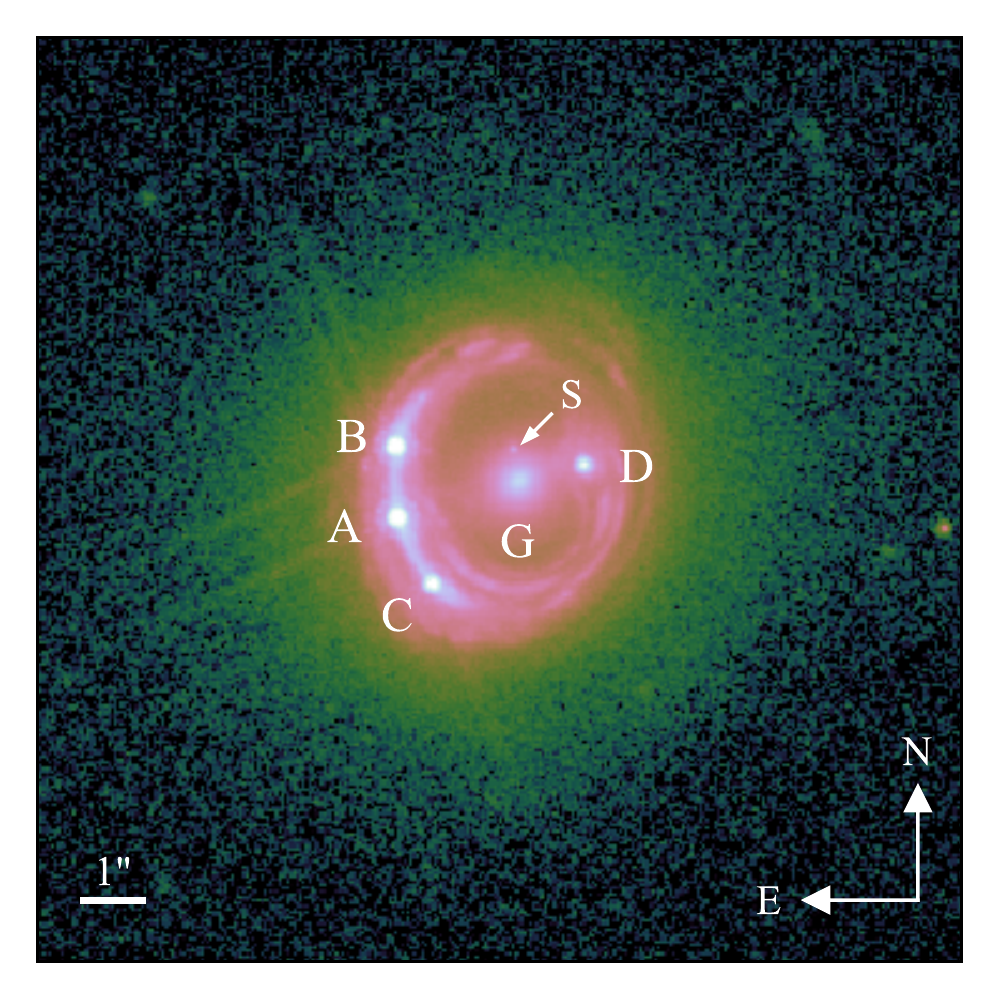}
	\caption{\label{fig:cutout}
	HST/ACS image of \lensname\ in the F814W band. \ajstwo{The four quasar images are labeled with A, B, C, and D. The central deflector is marked with G, of which we are measuring the spatially resolved velocity dispersion. An arrow points to the nearby satellite S, \ajsfou{which we mask out in the velocity dispersion measurement.}. The North and East directions and 1\" scale are also illustrated.} 
	}
\end{figure}

\subsection{Description of lens system} \label{sec:lens_description}

The quadruply imaged quasar lens system \lensname\ was discovered by \citet{Sluse03}.  \ajstwo{The deflector in this system is an elliptical galaxy with redshift $z_{\rm d} = 0.295$, and the source redshift is $z_{\rm s} = 0.657$ \citep{Sluse03}.} Due to its low redshifts, the system is relatively bright and large in angular size. The Einstein ring in this system contains intricate features, providing a wealth of information to constrain the lens mass model (see \figref{fig:cutout}). Due to its early discovery and information-rich features, this system is one of the most studied lensed quasar systems. The time delays for this system were measured by \citet{Tewes13}. \citet{Suyu13, Suyu14} performed cosmographic analyses of this system. These authors combined simply parametrized lens models based on the high-resolution imaging from the HST's \ajsthr{Advanced Camera for Surveys (ACS) instrument} (HST-GO 9744; PI: Kochanek), the measured time delays, single-aperture velocity dispersion, and external convergence estimate to infer $H_0 = 80.0_{-4.7}^{+4.5}$ \hunit. However, such simply parameterized lens models implicitly break the MSD. \citet{Birrer16} performed an independent mass modeling of this system while marginalizing the MSD with a prior on the source size. These authors found the prior choice on the anisotropy in the dynamical modeling to be the dominant systematic in inferring \Ho.

\subsection{KCWI Spectroscopy} \label{sec:kcwi_spectra}

We obtained integral field unit (IFU) spectroscopy of \lensname\ on 16 May and 7 June 2021 with the KCWI instrument on the Keck Observatory \citep[][]{Morrissey12, Morrissey18}. We chose KCWI with the small IFU slicer and the low-resolution blue grating (BL) with a field-of-view (FoV) of $8\farcs4 \times 20\farcs4$. The spectral resolution is $R\approx3600$, corresponding to an instrumental dispersion $\sigma_{\rm inst} \sim 35$ \kmps. \ajsthr{The reciprocal dispersion is $0.5\ \AA$ per pixel.} The observed wavelength range 3600--5600 \AA~ covers the Ca H\&K lines with wavelengths $\lambda\lambda3933$, $3968$ \AA\ at the redshift of the lens galaxy ($z_{\rm d} = 0.295$). We primarily use these lines to determine the stellar velocity dispersion. The redshifted 4304 \AA\ G-band is beyond the observed range, so it is not accessible with the KCWI for the \lensname\ system. 

\ajsthr{We aligned the FoV's longer side with the North direction (i.e., $\textrm{PA}=0\degr$) and dithered the individual exposures by $9\arcsec$ along the North-South direction. As the extent of the \lensname\ system is smaller than the FoV, each exposure contained the entire lens system within the FoV. In different exposures, the lens system occupied the upper or lower portion of the FoV. 
\ajsfou{Thus, the sky in an exposure with the system occupying the upper portion can be subtracted using another exposure with the system occupying the lower portion, and vice versa.}} We obtained six exposures with a total integration time of 10,560 s on 16 May and three with a total integration time of 5,400 s on 7 June. Therefore, the total exposure time is $t_{\textrm{exp}} = 15,960$ s. \ajstwo{The airmass ranged from 1.2 to 1.48 over the integrating period.}

\subsection{Data Reduction} \label{sec:reduction}

We use the official \textsc{Python}-based data reduction pipeline\footnote{developed by Luca Rizzi, Don Neill, Max Brodheim; \url{https://kcwi-drp.readthedocs.io/}} (DRP) to reduce our data. The DRP converts \ajstwo{the 2D raw data captured on the detector into a 3D datacube}. It performs geometry correction, \ajstwo{differential atmospheric refraction} correction, and wavelength calibration and produces a final standard-star-calibrated 3D datacube for each exposure. \ajsthr{The calibration with the standard star corrects for instrumental response and scales the data to flux units \citep{Morrissey18}.} We use the final output file with the suffix ``\texttt{\_icubes}'' for further analysis.

\ajsthr{We stack the dithered datacubes through drizzling \citep{Fruchter02}. Since the exposures are obtained on different dates, the world coordinate system information is not accurate enough to determine the relative positions of the dithered exposures. We follow \citet{Chen21b} to determine the relative positions by simultaneously fitting the point spread function (PSF) to the four quasar image positions. To perform the drizzling on the datacubes, we repurpose the drizzling routine of the DRP for OSIRIS, another IFU spectrograph on the Keck Observatory\footnote{\url{https://github.com/Keck-DataReductionPipelines/OsirisDRP}}. For the drizzling process, we set $\texttt{pixfrac}=0.7$ as recommended to reduce correlated uncertainties between the drizzled pixels \citep{Avila15b}. We calculate the drizzled weight image and ensure that the \ajstwo{ratio of} RMS/median $<0.2$ in the region of interest so that the trade-off is balanced between improving the image resolution and increasing the background noise \citep{Gonzaga12}. The rectangular pixel size $0\farcs1457\arcsec \times 0\farcs3395$ of the KCWI is kept the same in the drizzled output. We transform the datacube to have square pixels of size $0\farcs1457 \times 0\farcs1457$ through resampling while conserving the total flux. We converted the pixels into square sizes for the convenience of Voronoi binning the spectra using the software \textsc{vorbin} as described in \secref{sec:kinematics_measurement}.}

\begin{figure*}
    \centering
    \includegraphics[width=\textwidth]{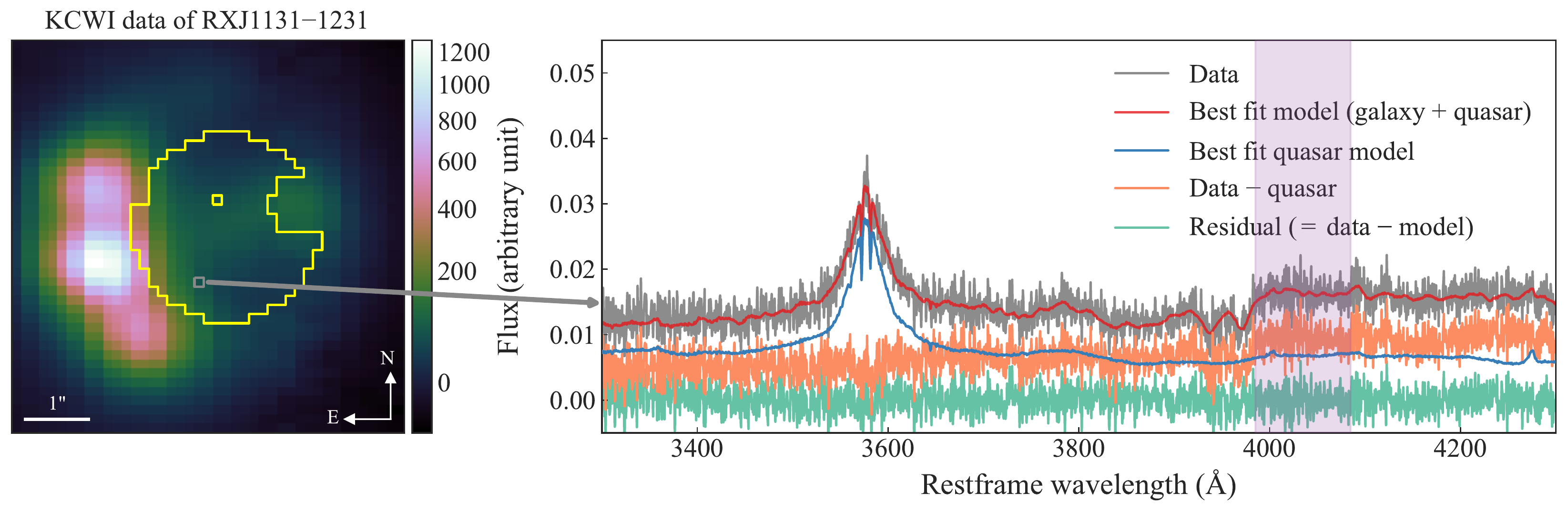}
    \caption{\label{fig:kcwi_datacube}
    \textbf{Left}: 2D representation (median-collapsed) of the 3D KCWI datacube for \lensname. The yellow contour traces the region with $1\farcs5$ radial extent from the center selected for stellar kinematic measurement. A circular region with $0\farcs5$ radius around image D and the spaxel containing the satellite S are excluded from this selected region. All the individual spaxels within this region have continuum $S/N > 1.4\ \AA^{-1}$ for the lens galaxy's light within 3985--4085 $\AA$ (the purple shaded range in the right panel). \textbf{Right}: The spectra (grey) from an example pixel (grey box in the left panel) and the estimate of the signal from the lens galaxy's spectra (orange) after removing the contribution from the quasar light (blue). The full model of the spectra is presented with the red line, and the model's residual is plotted in emerald color. The vertical purple shaded region marks where we compute the continuum $S/N$.}
\end{figure*}

\ajsthr{We directly estimate the PSF from the observed data. We produce a 2D image from the datacube by summing along the wavelength axis (see \figref{fig:kcwi_datacube}). We create a model for this KCWI image using a high-resolution template from the HST imaging (\figref{fig:cutout}) that has a pixel size $0\farcs05$ and PSF full width at half maximum (FWHM) $0\farcs10$. In the model, the template is convolved with a Gaussian PSF with a free FWHM parameter, and the positioning of the template on the KCWI image grid is fitted with two additional free parameters. By optimizing the model, we estimate that the PSF FWHM is 0\farcs96.}

\begin{figure}
	\includegraphics[width=\columnwidth]{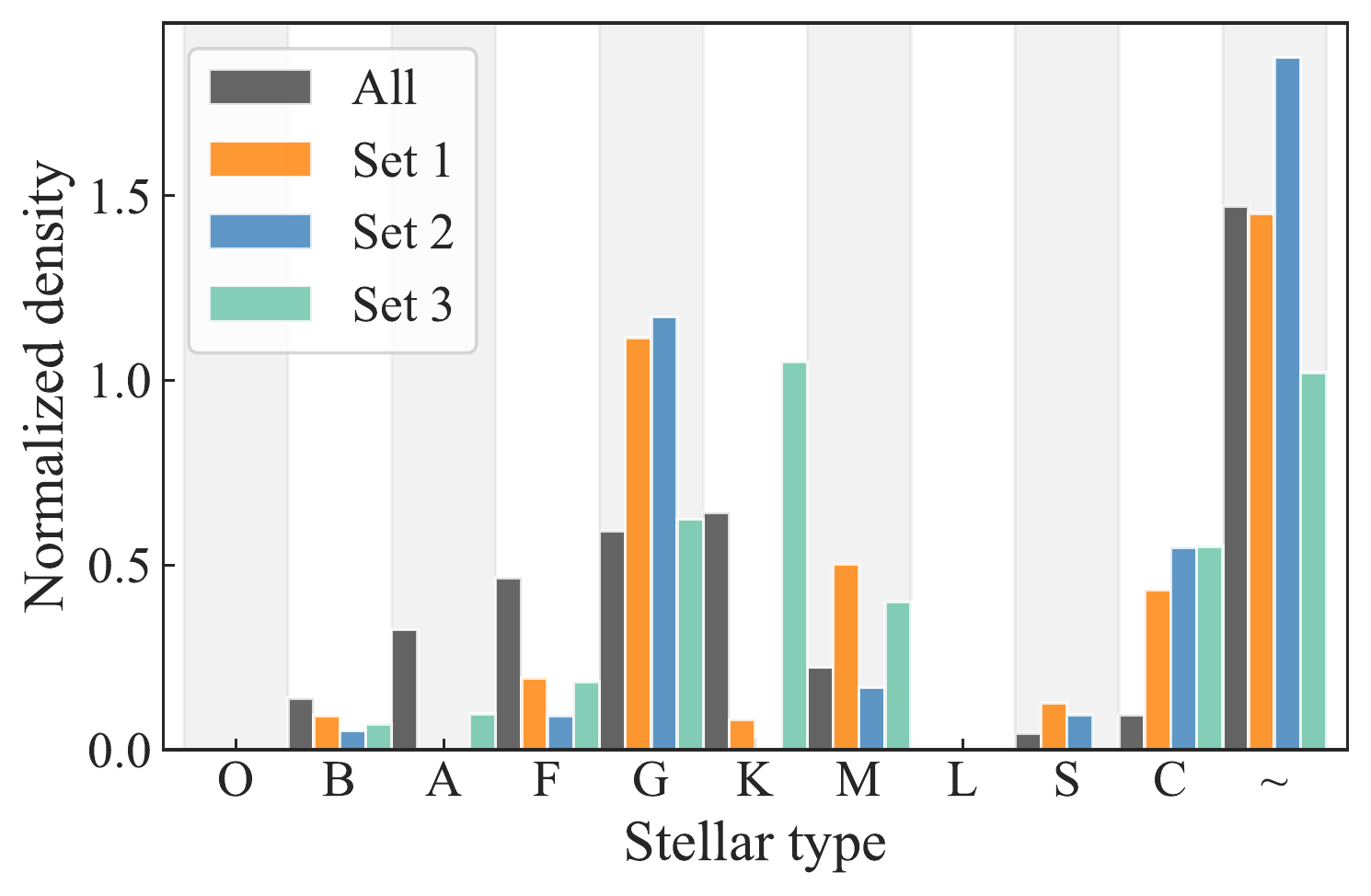}
	\caption{\label{fig:stellar_types}
	\ajsthr{Distribution of the stellar spectral types in the XSL according to the \textsc{Simbad} database. Unspecified stars are grouped in the `{$\sim$}' class. The dark grey color represents the full library of 628 stars. Set 1 (orange) refers to the 39 stars selected by \textsc{pPXF} out of the full library to construct an optimal template 1. Set 2 (blue) refers to 32 stars selected from a random half of the full library and set 3 (emerald) refers to 33 stars selected from the other half. Sets 2 and 3 have 15 and 17 stars, respectively, in common with Set 1. The alternating light grey and white vertical regions divide the spectral classes for easier visualization.}
	}
\end{figure}

\section{Kinematics maps} \label{sec:kinematics}

This section describes our procedure to obtain the final kinematics map. We use the \textsc{pPXF} package\footnote{\url{https://pypi.org/project/ppxf/}} to fit the spectra with a library of stellar templates and extract the velocity dispersion \citep{Cappellari17, Cappellari22}.  In \secref{sec:stellar_templates}, we describe the stellar templates used for the analysis.  In \secref{sec:kinematics_measurement}, we present the measurement of the spatially-resolved kinematics map of the lens galaxy. In \secref{sec:systematics}, we test the systematics of the velocity dispersion measurement.

\subsection{Library of Stellar Templates} \label{sec:stellar_templates}

The popularly used template libraries Medium-resolution Isaac Newton Telescope library of empirical spectra \citep[MILES;][]{Sanchez-Blazquez06} and INDO-US templates \citep{Valdes04} are both too low resolution to fit our datasets. The KCWI's instrumental resolution of $R \approx 3600$ leads to $\sigma_{\rm inst} \sim 35$ \kmps\ \ajsthr{for a Gaussian line spread function (LSF)}\footnote{\ajsthr{We quantitatively verified that the shape of the instrumental LSF is Gaussian \citep[\textit{cf.} Figure 28 of][]{Morrissey18}. Thus, the treatment of the instrumental LSF in \textsc{pPXF} is self-consistent and avoids any systematic bias due to inconsistent definitions of the LSF's FWHM \citep{Robertson13}.}}. MILES has a resolution of $\sigma_{\rm template} \sim 64$ \kmps\  (i.e., $R \sim 2000$), and the INDO-US templates have an approximately constant-wavelength resolution of 1.2 \AA, which corresponds to $\sigma_{\rm template}=39$ \kmps\ over the Ca H\&K wavelength range. Therefore, we choose the X-shooter Spectral Library (XSL), which contains 628 stars covering three segments, including UVB, Vis, and NIR bands \citep{Gonneau20}. As our data cover the rest-frame blue/UV range, we only use the UVB segment to fit the data, where its resolution is $R \sim 9700$ and $\sigma_{\rm template} \sim 13$ \kmps.

\begin{figure*}
    \centering
    \includegraphics[width=\textwidth]{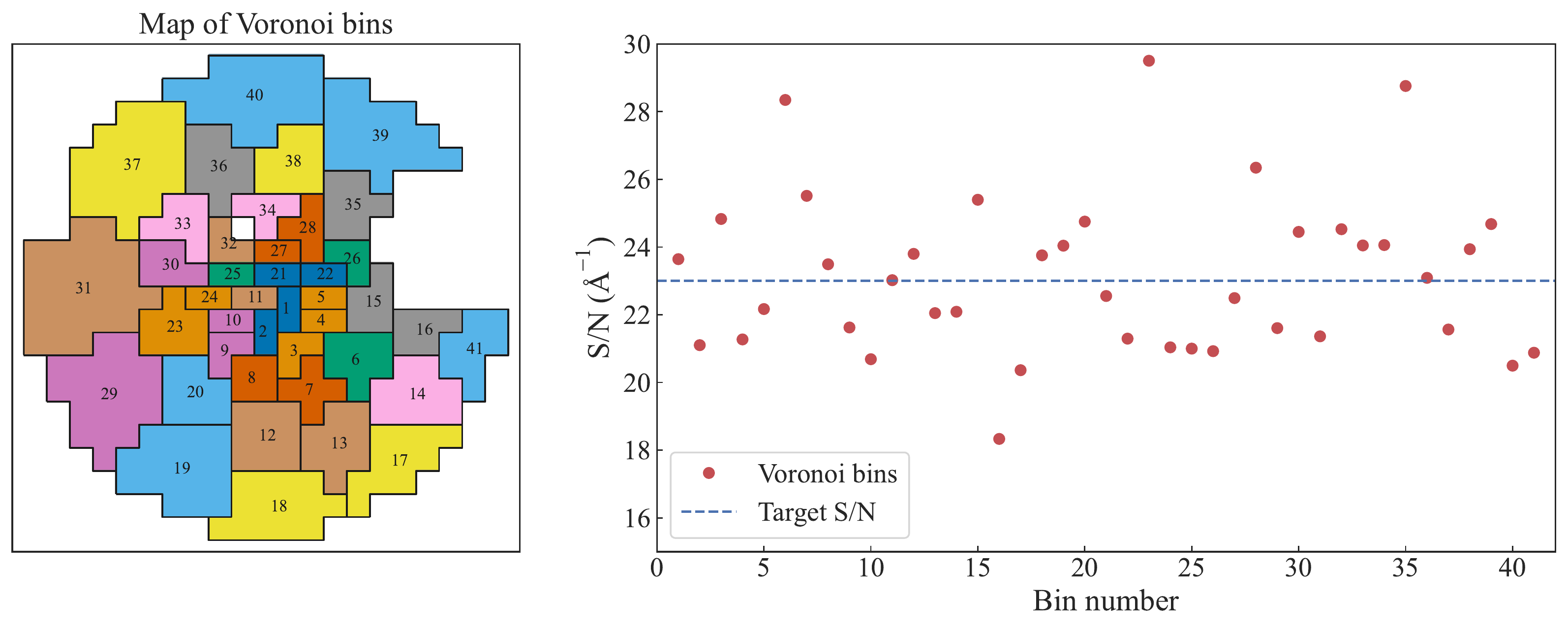}
    \caption{\label{fig:voronoi}
    \textbf{Left:} Voronoi binning of the selected spaxels within $1\farcs5$ from the galaxy center that avoid lensed arcs, quasar images, and the satellite galaxy S. \ajstwo{The different colors illustrate the regions for each Voronoi bin in a cartographic manner for easier visualization, with the bin number specified within each bin.} We perform the binning with a target $S/N \approx23\ \AA^{-1}$ for each bin, which results in 41 bins in total. \textbf{Right:} Resultant $S/N$ for each Voronoi bin (red points).}
\end{figure*}

\begin{figure*}
    \centering
    \includegraphics[width=\textwidth]{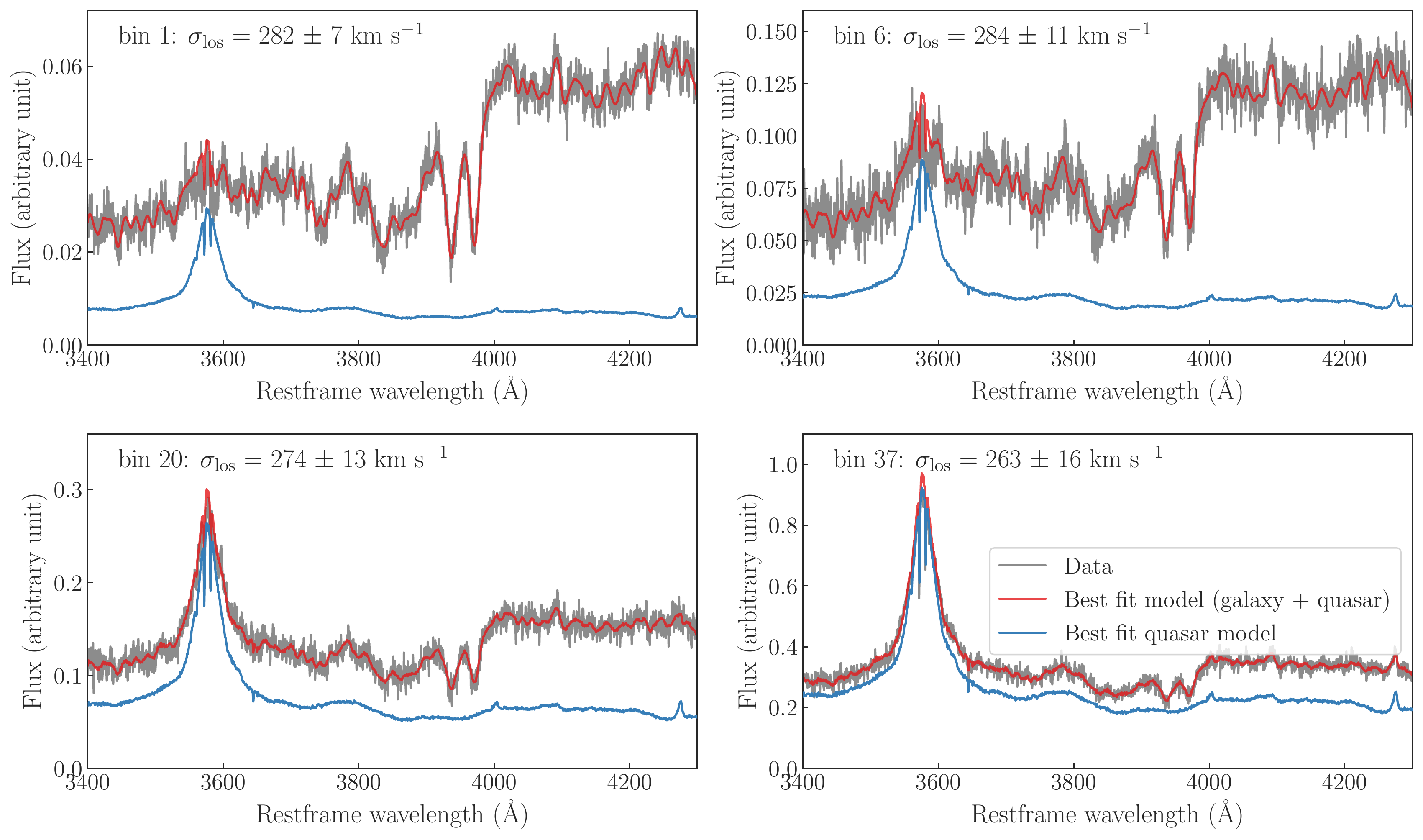}
    \caption{\textsc{pPXF} fitting to the spectra from four examples of Voronoi bins. The bin number and the measured velocity dispersion for the corresponding bin are specified in each panel. The grey line presents the full spectra, the red line traces the best-fit model, and the blue line shows the quasar component in the best-fit model.}
    \label{fig:vd_demo}
\end{figure*}

\subsection{Measuring the velocity dispersion} \label{sec:kinematics_measurement}

\ajsfou{We choose a cutout centred on the lens system with $6\farcs235 \times 6\farcs235$ (43 pixels $\times$ 43 pixels) to initiate the analysis (see \figref{fig:kcwi_datacube}). We estimate the lens galaxy light's signal-to-noise ratio ($S/N$) in each spatial pixel (hereafter, spaxel) within this initial cutout. We then select a region with sufficient $S/N$ from the lens galaxy and relatively low quasar contamination for measuring the velocity dispersion (the yellow contour in \figref{fig:kcwi_datacube}'s left panel). We perform Voronoi binning within this selected region to preserve the maximal spatial resolution and reduce the bias in the lower-$S/N$ region \citep{Cappellari03}. We elaborate on these steps below.}

To estimate the lens galaxy's $S/N$ in each spaxel, we first simultaneously fit the quasar and the lens galaxy in each spaxel to calculate the signal from each of them. \ajsfou{We perform this fitting within the wavelength range 3400--4300 \AA.} As the four quasar images surround the lens galaxy, each spaxel receives a different contribution from the quasar light. We take spectra at the central spaxel of image A as the quasar template, ignoring the lens galaxy's small contribution. \ajsthr{Later in \secref{sec:systematics}, \ajsfiv{we also choose the quasar template from images B and C to account for the associated systematic uncertainty, i.e., the potential impact of chromatic microlensing that may change the contrast between the line and the continuum \citep[e.g.,][]{Sluse07}.}}

\ajsthr{We determine a single optimal template spectrum for the lens galaxy template. For this purpose, we binned the spectra from spaxels within a circular region of radius $0\farcs5$ centered on the lens galaxy and fit it with \textsc{pPXF} using the 628 stellar templates from the XSL and the quasar template. We also include a Legendre polynomial of degree 3 as a component in the fitting to account for any residual gradient in the continuum. \textsc{pPXF} chooses 39 of the stellar templates and builds the optimal template by taking a weighted linear combination of them. See \figref{fig:stellar_types} for the weighted distribution of spectral types of the full template library and that of the 39 stars selected by \textsc{pPXF}. Among those stars in the XSL with stellar classes specified by the \textsc{Simbad} database \citep{Wenger00}, G-type stars are selected with the highest total weight, consistent with the fact that massive elliptical galaxy spectra are dominated by G and K-type stars. In the \textsc{pPXF} fitting procedure, the stellar templates are broadened, corresponding to a freely varying velocity dispersion, but the velocity dispersion does not broaden the quasar template.}

\ajsthr{Once the optimal galaxy template is constructed, we use this template and the quasar template to fit the spectrum of each spaxel individually. \ajsfou{We use this optimal template to fit the galaxy spectra in individual spaxels instead of the full template library to avoid large spurious fluctuations in the measured velocity dispersion from spaxel to spaxel.} We show the decomposition of the spectra from one example spaxel into different components after fitting with \textsc{pPXF} in \figref{fig:kcwi_datacube}. We calculate the signal of the lens galaxy's spectrum in each spaxel by subtracting the modeled quasar component from the observed spectra.} \ajsfou{The noise is estimated by adding in quadrature the Poisson noise of the total signal and the background noise estimated from an empty patch of the sky. \ajsfiv{The noise values are multiplied by $\sqrt{2}$ to account for the fact that the square pixels are created from the rectangular pixels about double the size through resampling.}} We estimate the $S/N$ using the restframe wavelength range 3985--4085 \AA, slightly above the Ca H\&K absorption lines in wavelength (see the purple shaded region in \figref{fig:kcwi_datacube}).

To perform Voronoi binning before the velocity dispersion measurement, we select the spaxels within a radius of \ajsthr{$1\farcs5$\footnote{For reference, $1\farcs5$ corresponds to 6.6 kpc at $z_{\rm d} = 0.295$ for a fiducial flat \lcdm\ cosmology with $\ho = 70$ \hunit\ and $\Omega_{\rm m} = 0.3$.} from the lens galaxy center that avoid the brightest spaxels containing images A, B, and C and the lensed arcs. We also exclude a circular region around image D with radius $0\farcs5$. To avoid any potential bias due to contamination from the satellite galaxy S, we exclude the spaxel at its position \citep[$\Delta{\rm RA}=0\farcs09$, $\Delta{\rm Dec}= 0\farcs54$ from the galaxy center,][]{Suyu13}. We also exclude pixels with $S/N < 1\ \AA^{-1}$. In the end, the spaxels within the selected region have $S/N >1.4\  \AA^{-1}$ (see \figref{fig:kcwi_datacube} for the selected region).} We perform Voronoi binning using \textsc{vorbin}\footnote{\url{https://pypi.org/project/vorbin/}} given the estimated $S/N$ values for each spaxel. In \figref{fig:voronoi}, we show the 41 Voronoi bins obtained by setting the target $S/N \approx 23\ \AA^{-1}$ for each bin. This target $S/N$ was chosen so that the resultant $S/N \gtrsim 20\ \AA^{-1}$ for each bin, which is standard practice (\figref{fig:voronoi}, only bin 16 has $S/N \approx 18\ \AA^{-1}$).

For each Voronoi bin, we measure the velocity dispersion by fitting the binned spectra using \textsc{pPXF} using the optimal galaxy template described above, the quasar template, and the additive Legendre polynomial to model any slight gradient in the population. A few examples of \textsc{pPXF} fit of the binned spectra are shown in \figref{fig:vd_demo}.

\subsection{Estimation of systematic uncertainty} \label{sec:systematics}

\begin{figure*}
	\includegraphics[width=\textwidth]{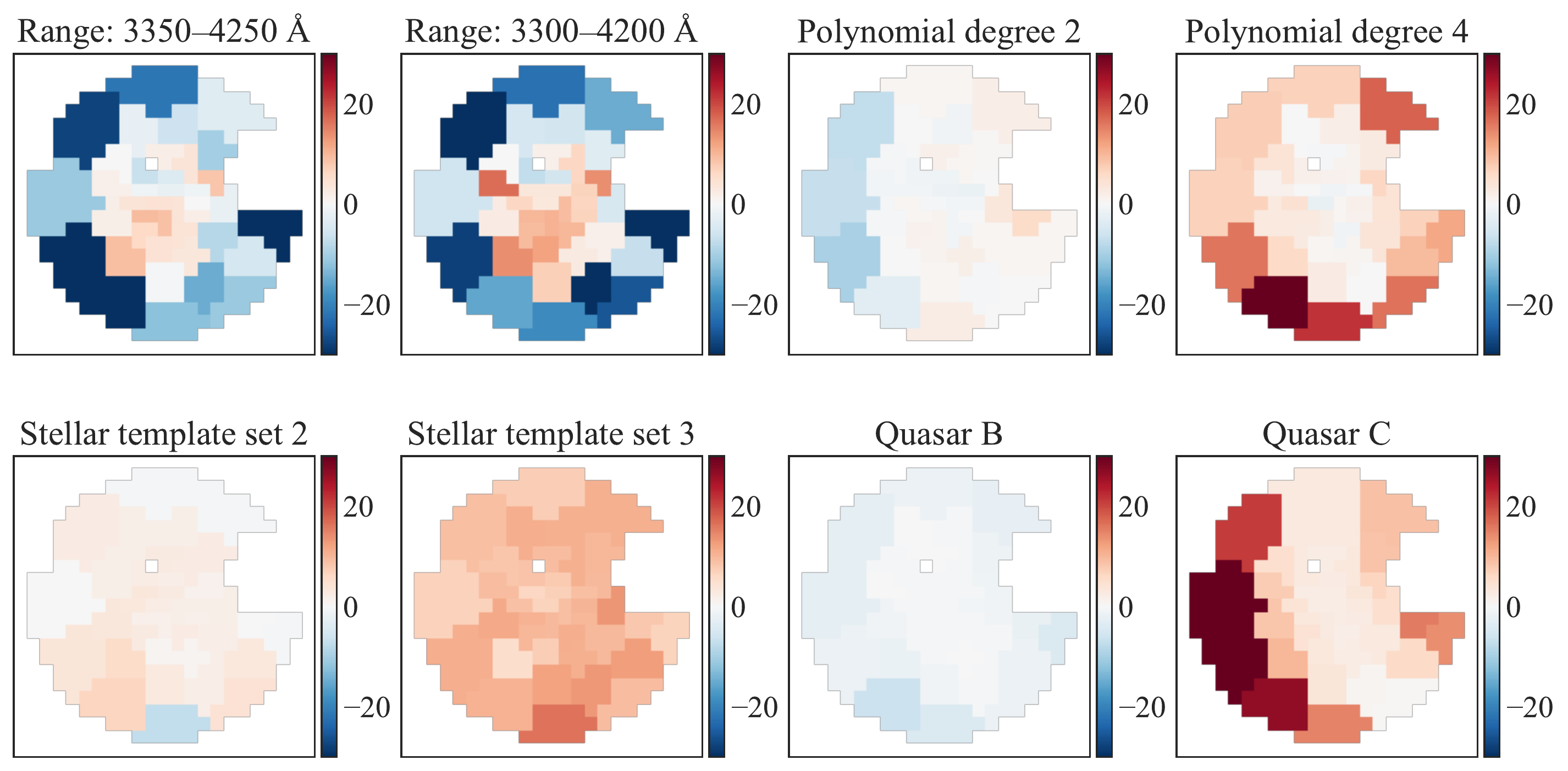}
	\caption{\label{fig:absolute_systematic}
	Absolute difference in \kmps\  between the extracted velocity dispersion from two setups that differ by one setting. The baseline setup has the range: 3400--4300 \AA, polynomial degree: 3, stellar template set 1, and quasar template from image A. The different setting for each case is specified at the top of each panel.
	}
\end{figure*}

\begin{figure*}
	\includegraphics[width=\textwidth]{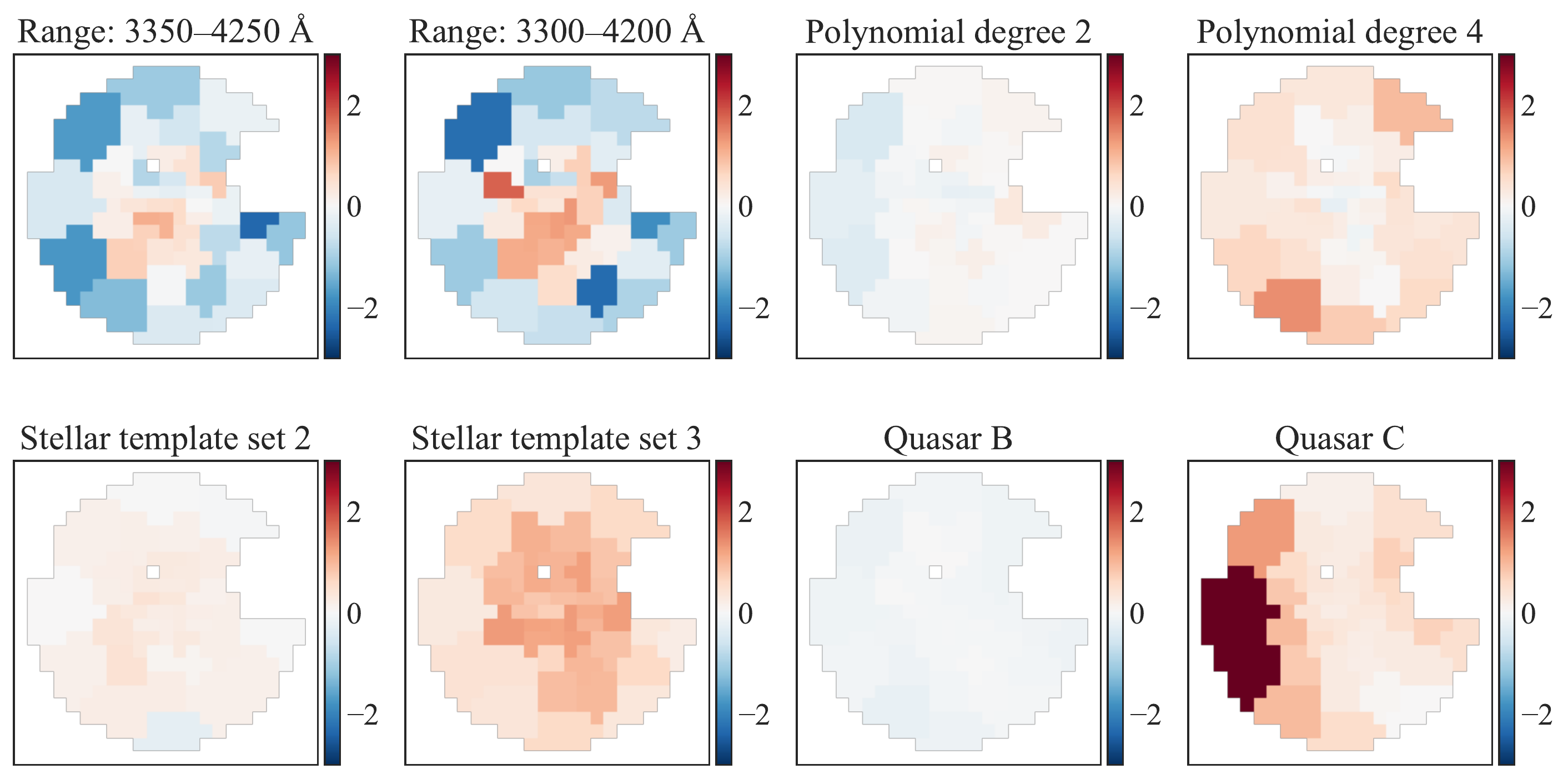}
	\caption{\label{fig:normalized_systematic}
	Same as \figref{fig:absolute_systematic}, but the difference is normalized by the statistical uncertainty of the baseline setup.
	}
\end{figure*}

To estimate the systematic uncertainties in the velocity dispersion measurement, we consider a range of plausible choices in the extraction procedure: the degrees of the additive Legendre polynomial used to correct the template continuum shape between 2 to 4; the quasar template obtained from images A, B, and C; the fitted wavelength range chosen from 3300--4200 \AA, 3350--4250 \AA, and 3400--4300 \AA; and three sets of template spectra used in the fitting. \ajsthr{The first set of template spectra contains the complete XSL of 628 stars. The second set contains half of the entire sample that is randomly selected, and the third set contains the other half. The numbers of stars selected by \textsc{pPXF} in the three sets are 39, 32, and 33, respectively. Sets 2 and 3 have 15 and 17 stars, respectively, in common with Set 1. \figref{fig:stellar_types} shows the distribution of spectral types in all three sets and the entire library.} We do not take the quasar template from image D as it is much fainter than the other images, and thus the galaxy contribution in the brightest spaxel on image D is non-negligible. Taking a combination of all of these choices yields 81 different setups. \ajsthr{We illustrate the shift in the extracted velocity dispersion maps for one change of setting at a time in Figures \ref{fig:absolute_systematic} and \ref{fig:normalized_systematic}.} 

\begin{figure}
    \centering
    \includegraphics[width=0.5\textwidth]{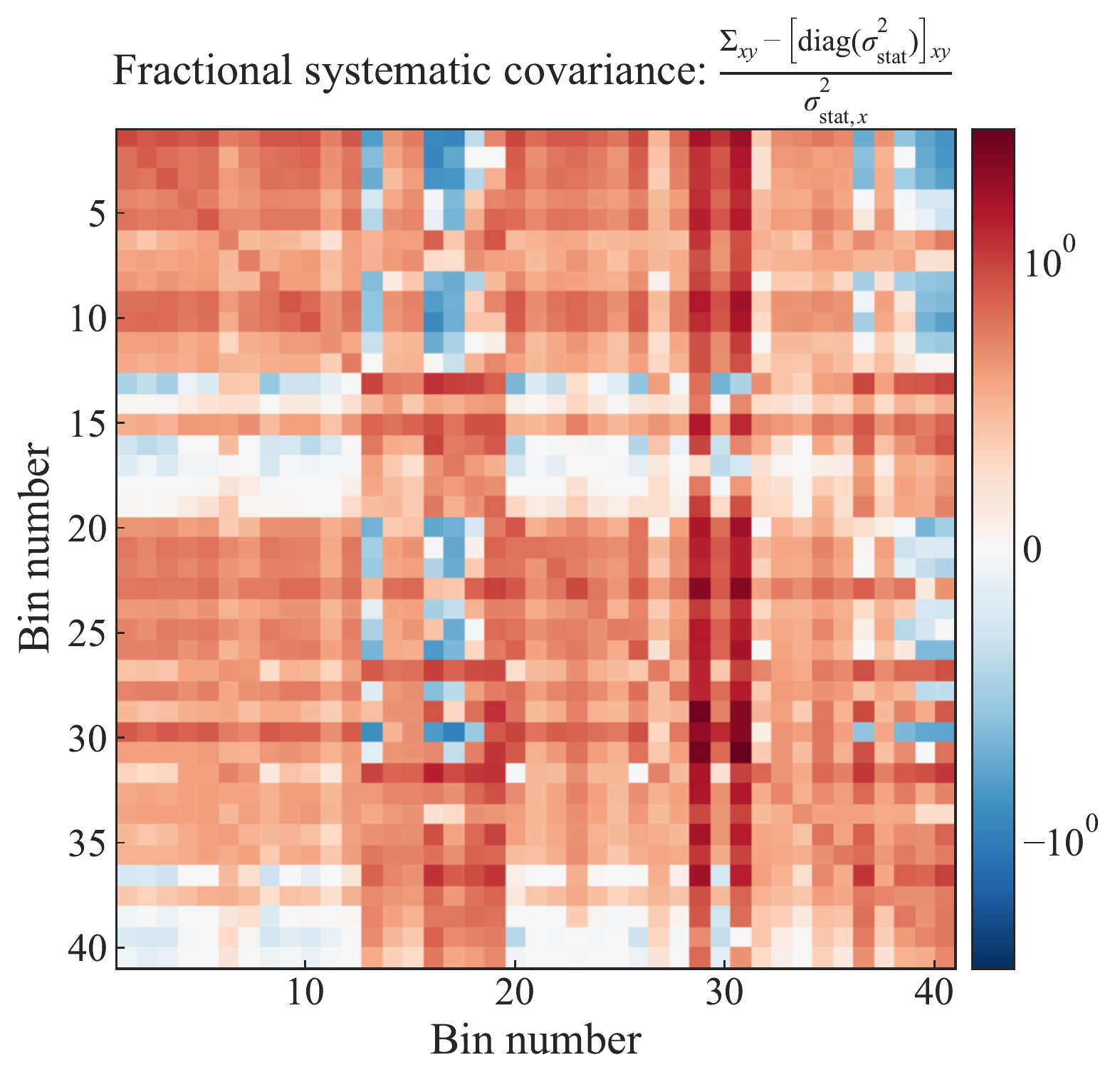}
    \caption{ \label{fig:covariance_matrix}
    \ajsfiv{Illustration of the systematic covariance relative to the statistical covariance. $\Sigma$ is the variance-covariance matrix of the Voronoi-binned velocity dispersions (with target $S/N \approx 23\ \AA^{-1}$ for each bin),  $\sigma_{\rm stat, \it x}$ is the statistical uncertainty in bin number $x$ from our fiducial setup, and ${\rm diag}(\sigma_{\rm stat})$ is a diagonal matrix. Note that we assume no covariance in the statistical uncertainty from each setup for kinematic measurement. Thus the off-diagonal terms in the variance-covariance matrix purely represent the systematic covariance. Most diagonal terms are $<1$ (with a median of 0.47), showing that the systematic variances are subdominant to the statistical variances except for bins 29 and 31. These bins are close to images A and C. Thus, they are largely susceptible to the choice of the quasar template, as seen in Figures \ref{fig:absolute_systematic} and \ref{fig:normalized_systematic}.}
    }
\end{figure}

We estimate the variance-covariance matrix of the binned velocity dispersions from these 81 setups. \ajsfou{To do this, we generate 1,000 random realizations of the measured velocity dispersion map for each of the 81 setups using the corresponding statistical uncertainty. We create the variance-covariance matrix from the 81,000 realizations combined from all the setups. In this way, the diagonal terms of the variance-covariance matrix encode the total variance from systematic and statistical uncertainties, and the off-diagonal terms encode the systematic covariances. For example, if all 81 setups hypothetically provided the same velocity dispersion map and uncertainty, then the off-diagonal terms would be zero, and the diagonal terms would reflect only the statistical uncertainties.} \ajsfiv{We show the systematic variance-covariance relative to the statistical variance in \figref{fig:covariance_matrix}. The systematic variance is subdominant relative to the statistical variance (with a median of 0.47 of the ratio between systematic and statistical covariances along the diagonal) except for bins 29 and 31. These two bins are closest to quasar images A and C, and thus largely susceptible to the choice of quasar template (see Figures \ref{fig:absolute_systematic} and \ref{fig:normalized_systematic}.)}

\begin{figure}
    \centering
    \includegraphics[width=0.5\textwidth]{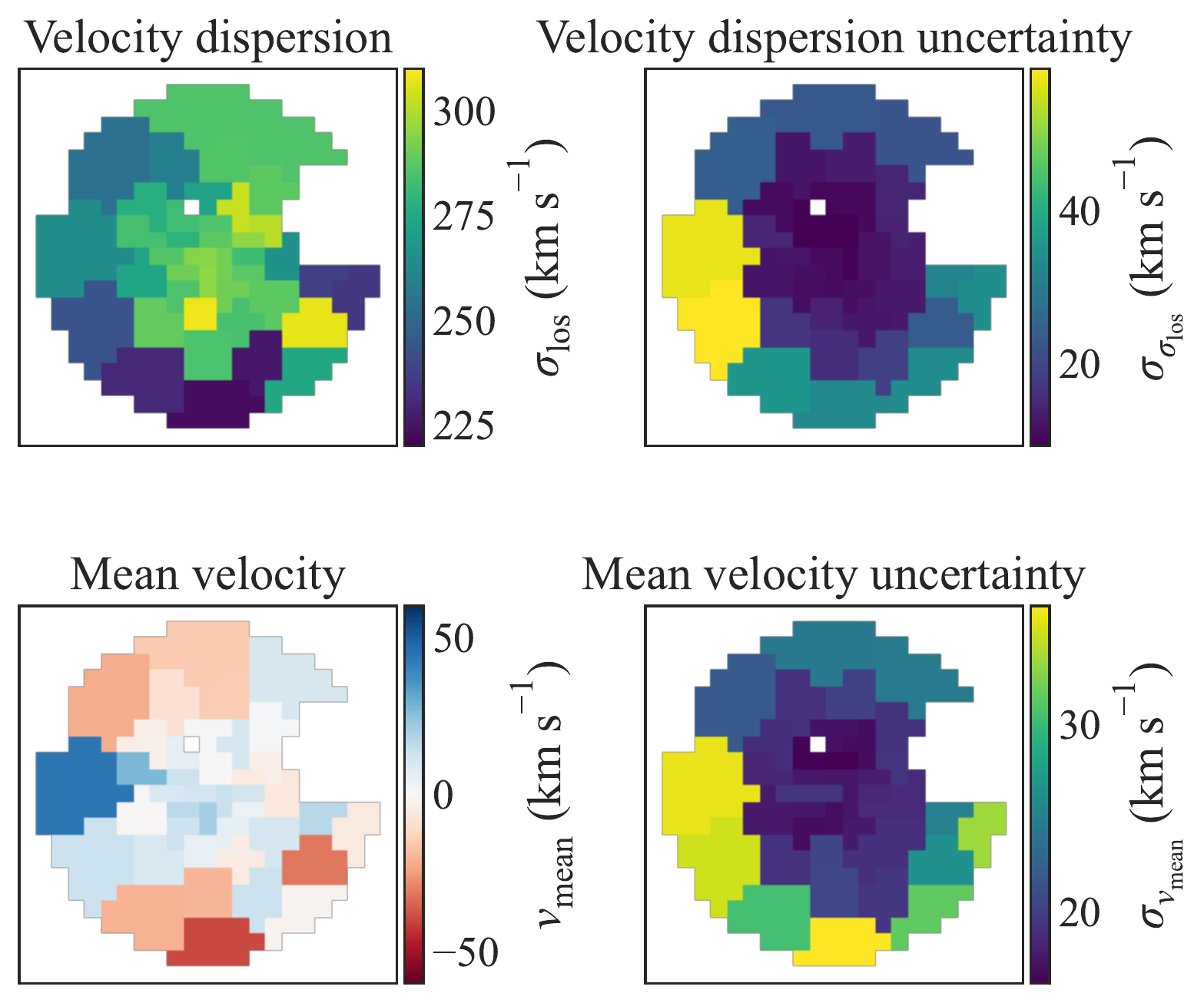}
    \caption{Maps of extracted velocity dispersion (top row) and mean velocity (bottom row) in Voronoi bins along with the corresponding uncertainties (right column). The Voronoi binning was tuned to achieve $S/N \approx 23 \ \AA^{-1}$ for each bin. The illustrated maps (left column) correspond to the average values after combining 81 model setups, and the uncertainty maps correspond to the square root of the diagonal of the variance-covariance matrices. A systematic velocity of 182 \kmps\ was subtracted from the mean velocity map.}
    \label{fig:kinematic_maps}
\end{figure}

We show the velocity dispersion and mean velocity maps averaged over the 81 setups in \figref{fig:kinematic_maps}. We estimate a systematic velocity of 182 \kmps\ using the \textsc{pafit}\footnote{\url{https://pypi.org/project/pafit/}} software program \citep{Krajnovic06} and subtract it from the mean velocity map. \ajstwo{The systematic velocity is the result of a slight deviation in the true redshift from the fiducial value.} The mean velocity map does not show any significant evidence of ordered rotation above the systematic and statistical noise levels. Thus it is consistent with the lens galaxy being a slow rotator. We \ajsfiv{use this} systematic-averaged velocity dispersion map and \ajsfiv{the variance-covariance matrix estimated above when computing} the likelihood function for dynamical modeling in \secref{sec:dynamic_models}.

\begin{figure}
	\includegraphics[width=\columnwidth]{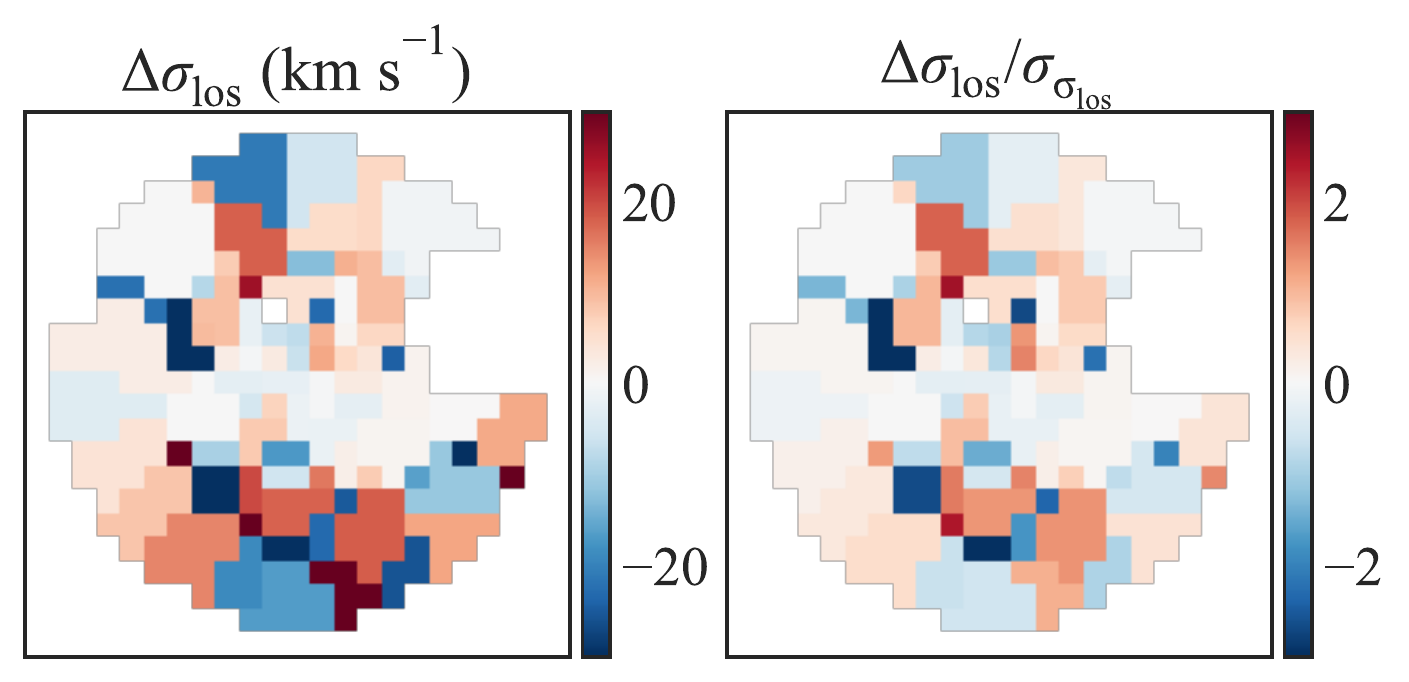}
	\caption{\label{fig:binning_systematic}
	Absolute (left) and uncertainty-normalized (right) difference in the extracted velocity dispersion between two Voronoi binning schemes. The two binning schemes are obtained by setting the target $S/N$ to 23 $\AA^{-1}$ and 28 $\AA^{-1}$ for each bin. We take the case with target $S/N \approx 23 \ \AA^{-1}$ for each bin as the baseline in our analysis.
	}
\end{figure}

To test the impact of our choice for the Voronoi binning scheme, we adopt an alternative target $S/N \approx 28\ \AA^{-1}$ for each bin, which results in 27 bins. We similarly produce another set of 81 model setups in this binning scheme and produce the variance-covariance matrix for these binned velocity dispersions. We test the systematic impact of this different binning scheme on the cosmological measurement later in \secref{sec:systematic_checks}. \ajsthr{We show the difference in the extracted kinematics between the two binning schemes in \figref{fig:binning_systematic}.}

\section{Overview of lens and dynamical modeling} \label{sec:theory}

This section reviews the theoretical formalism of lens and dynamical modeling.

\subsection{Lensing observables and modeling}

We briefly review the strong lensing formalism in the context of time-delay cosmography in \secref{sec:lensing_formalism}, describe the mass-sheet transform (MST) in \secref{sec:mst}, and explain the internal and external components of the MST in \secref{sec:internal_mst}.

\subsubsection{Strong lensing formalism} \label{sec:lensing_formalism}

In the thin lens approximation applicable in this case, lensing observables are described using the surface mass density $\Sigma(R)$ projected from the 3D mass density distribution $\rho(r)$ in the lens galaxy. Formally, the lensing observables depend on the dimensionless convergence defined as
\begin{equation}
	\kappa (\btheta) \equiv \frac{\Sigma (\btheta)}{\Sigma_{\rm cr}},
\end{equation}
which is the surface mass density normalized by the critical density
\begin{equation}
	\Sigma_{\rm cr} \equiv \frac{c^2 \Ds}{4 \uppi G \Dd \Dds }.
\end{equation}
Here, $c$ is the speed of light, $G$ is the gravitational constant, $\Ds$ is the angular diameter distance between the observer and the source, $\Dd$ is the angular diameter distance between the observer and the lens galaxy, and $\Dds$ is the angular diameter distance between the lens galaxy and the source. The \ajstwo{on-sky} deflection angle $\balpha(\btheta)$ relates to the convergence as 
\begin{equation}
	\kappa (\btheta) = \frac{1}{2} \nabla \cdot \balpha (\btheta).
\end{equation}
The time delay between two quasar images labeled A and B is given by
\begin{equation}
	 \begin{split}
        \Delta t_{\rm AB} &=  \frac{\Ddt}{c} \left[ \frac{(\btheta_{\rm A} - \bvarsigma)^2}{2} - \frac{(\btheta_{\rm B} - \bvarsigma)^2}{2} - \psi(\btheta_{\rm A}) + \psi(\btheta_{\rm B})  \right],
        %&= \frac{\Ddt}{c} \Delta \phi_{\rm XY}^{\rm eff}.
        \end{split}
\end{equation}
where \ajstwo{$\btheta_{\rm A}$ is the angular position of image A, $\bvarsigma$ is the source's angular position, $\psi(\btheta)$ is the lensing potential, and} the time-delay distance $\Ddt$ is defined as
\begin{equation}
	\Ddt \equiv ( 1+z_{\rm d}) \frac{\Dd \Ds }{\Dds} .
\end{equation}

\subsubsection{Description of the  MST} \label{sec:mst}

\ajsthr{The MST is a mathematical transform of the convergence profile that leaves invariant all the imaging observables, such as the image positions and the flux ratios \citep{Falco85, Schneider14}.} This transform scales the convergence and the unknown source position as
\begin{equation}
	\begin{aligned}
		&\kappa \to \kappa^\prime = \lambda_{\rm MST} \kappa + ( 1 - \lambda_{\rm MST}), \\
		&\varsigma \to \varsigma^\prime = \lambda_{\rm MST} \varsigma.
	\end{aligned}
\end{equation}
where $\lambda_{\rm MST}$ is the transformation parameter. The predicted time delay $\Delta t$ scales under the transform as
\begin{equation}
	\begin{aligned}
		& \Delta t \to \Delta t^\prime = \lambda_{\rm MST} \Delta t. \\
%		& \sigma_{\rm los, ap} \to \sigma_{\rm los, ap}^{\prime} = \sqrt{\lambda_{\rm MST}} \sigma_{\rm los,ap}.
	\end{aligned}
\end{equation}
Then, the inferred time-delay distance $\Ddt$ and the Hubble constant \Ho\ based on the observed time delays will change as 
\begin{equation}
	\begin{aligned}
		&\Ddt^\prime = \frac{\Ddt}{\lambda_{\rm MST}}, \\
		& H_0^{\prime} = \lambda_{\rm MST} H_0.
	\end{aligned}
\end{equation}
\ajsthr{However, the MST changes the predicted velocity dispersion, thus measuring it breaks the MSD. Notably, the MST also rescales the lensing magnifications. Thus, standardizable candles can also be used to break the MSD \citep{Bertin06, Birrer22} provided that microlensing and millilensing can be mitigated \citep[e.g.,][]{Yahalomi17, More17b, Foxley-Marrable18}.}

\subsubsection{Internal and external MST} \label{sec:internal_mst}

\ajsthr{We can express the ``true'' (i.e., physically present) lensing mass distribution as}
\begin{equation}
	\kappa_{\rm true} = \kappa_{\rm gal} + \kappa_{\rm ext},
\end{equation}
\ajsthr{where $\kappa_{\rm gal}$ is the mass distribution of the central lens galaxy (or galaxies) that is (are) considered in the lens modeling, and $\kappa_{\rm ext}$ is called the external convergence, which approximates the projected mass distribution of line-of-sight structures as a mass sheet. Since $\lim_{\theta \to \infty} \kappa_{\rm gal} = 0$ has to be satisfied, we find that $\lim_{\theta \to \infty} \kappa_{\rm true} = \kappa_{\rm ext}$, hence the interpretation of $\kappa_{\rm ext}$ as the lensing mass far from (or, ``external'' to) the central deflector(s).}

\ajsthr{All the lensing observables including imaging observables result from $\kappa_{\rm true}$. However, since only the central galaxies are usually considered in lens modeling with imaging observables, the lens model provides $\kappa_{\rm model}^\prime$ with $\lim_{\theta \to \infty} \kappa_{\rm model}^\prime = 0$. This $\kappa_{\rm model}^\prime$ is an MST of $\kappa_{\rm true}$ for $\lambda_{\rm MST} = 1/(1 - \kappa_{\rm ext})$ as
}
\begin{equation}
	\kappa_{\rm model}^{\prime} = \frac{\kappa_{\rm gal} + \kappa_{\rm ext}}{1-\kappa_{\rm ext}} + 1 - \frac{1}{1 + \kappa_{\rm ext}} = \frac{\kappa_{\rm gal}}{1 - \kappa_{\rm ext}}.
\end{equation}
\ajsthr{Lens mass models are usually described with simply parametrized models, such as the power law or a combination of the NFW profile and the observed stellar distribution. In that case, the assumption of a simple parametric form implicitly breaks the MSD. Therefore, the simply parametrized model $\kappa_{\rm model}$ can be expressed as another approximate MST of the $\kappa_{\rm model}^{\prime}$ as}
\begin{equation} \label{eq:approximate_mst}
	\kappa_{\rm model}^\prime \approx \lambda_{\rm int} \kappa_{\rm model} + (1 - \lambda_{\rm int}) \kappa_{\rm s} (\theta),
\end{equation}
\ajsthr{where $\lambda_{\rm int}$ is called  the internal MST parameter, and $\kappa_{\rm s}$ is a ``variable'' mass sheet with $\lim_{\theta \to \infty} \kappa_{\rm s} (\theta) = 0$ to ensure that both $\lim_{\theta \to \infty} \kappa_{\rm model}^{\prime} = 0$ and $\lim_{\theta \to \infty} \kappa_{\rm model} = 0$ are satisfied. However, for \equref{eq:approximate_mst} to be an approximate MST, the variable mass-sheet needs to satisfy $\kappa_{\rm s}(\theta) \simeq 1$ within the central region that lensing observables are sensitive to \citep[$\theta \lesssim 2\theta_{\rm E}$,][]{Schneider13}. This can be achieved with the formulation \citep{Blum20}}
\begin{equation} \label{eq:variable_mass_sheet}
	\kappa_{\rm s}(\theta) = \frac{\theta_{\rm s}^2}{\theta^2 + \theta_{\rm s}^2},
\end{equation}
\ajsthr{where $\theta_{\rm s} \gg \theta_{\rm E}$ is a scale radius where the variable mass-sheet smoothly transitions from $1 - \lambda_{\rm int}$ to 0. This approximate MST converges to the pure MST in the limit $\theta_{\rm s} \to \infty$.} \ajsthr{Thus, the actual mass distribution of the central deflector(s) relates to the modeled mass distribution as}
\begin{equation} \label{eq:complete_mst}
	\kappa_{\rm gal} \approx (1 - \kappa_{\rm ext}) \left[\lambda_{\rm int} \kappa_{\rm model} + (1 - \lambda_{\rm int}) \kappa_{\rm s} (\theta)\right].
\end{equation}

\ajsthr{The external convergence $\kappa_{\rm ext}$ can be estimated by using relative number counts of line-of-sight galaxies near the central deflector(s) \citep[e.g.,][]{Suyu10, Greene13, Rusu17, Buckley-Geer20}, or by using weak lensing of distant galaxies by the line-of-sight mass distribution \citep[e.g.,][]{Tihhonova18}. The measured velocity dispersion then constrains the internal MST parameter $\lambda_{\rm int}$ \citep{Birrer20,Yildirim21}.}

\subsection{Dynamical modeling}
\label{sec:rxj_model}

In this section, we describe the Jeans anisotropic multi-Gaussian-expansion (JAM) framework to model our dynamical observable, which is the spatially resolved stellar velocity dispersion measured in \secref{sec:kinematics}. The orbital motions of the stars, i.e., the distribution function $f(\bm{x}, \bm{v})$ of position $\bm{x}$ and velocity $\bm{v}$, in the galactic potential $\Phi$ is described by the steady-state collisionless Boltzmann equation  \citep[][Eq. 4-13b]{Binney87}
\begin{equation}
	\sum_{i=1}^3 \left( v_i \frac{\partial f}{\partial x_i} - \frac{\partial \Phi}{\partial x_i} \frac{\partial f}{\partial v_i} \right) = 0.
\end{equation}
We assume an axisymmetric case (i.e., $\partial \Phi/ \partial \phi = \partial f/ \partial \phi = 0$ with $\phi$ being the polar angle in the spherical coordinate system), a spherically aligned velocity ellipsoid, and the anisotropy for each Gaussian component in the multi-Gaussian expansion \citep[MGE;][]{Emsellem94, Cappellari02} to be spatially constant. Slow rotators such as the deflector galaxy in \lensname\ are in general expected to be weakly triaxial or oblate but never flat and instead quite close to spherical in their central parts \citep[e.g.,][]{Cappellari16}. For this reason, we expect the spherical alignment of the velocity ellipsoid of \textsc{jam}\textsubscript{sph} \citep{Cappellari20} to provide a better approximation to the galaxy dynamics than the cylindrical alignment \textsc{jam}\textsubscript{cyl} solution \citep{Cappellari08}. Then, the above equation gives two Jeans equations in spherical coordinates  \citep{Jeans22, Bacon83, deZeeuw96, Cappellari20}
\begin{equation} \label{eq:jeans}
	\begin{aligned}
		\frac{\partial \left(\zeta \moment{v_r^2} \right)}{\partial r} + \frac{(1 + \beta) \zeta \moment{v_r^2} -  \zeta \moment{v_\phi^2} }{r} &= -\zeta \frac{\partial \Phi}{\partial r}, \\
		(1 - \beta) \frac{\partial \left( \zeta \moment{v_r^2} \right) }{\partial \theta} + \frac{(1 - \beta) \zeta \moment{v_r^2} -  \zeta \moment{v_\phi^2} }{\tan \theta} &= -\zeta \frac{\partial \Phi }{\partial \theta},
	\end{aligned}
\end{equation}
where the following notations are used
\begin{equation}
	\begin{aligned}
		\zeta \moment{v_p v_q} &\equiv \int v_p v_q f \rmd^3 \bm{v}, \\
		\beta &\equiv 1 - \frac{\moment{v_\theta^2}}{\moment{v_r^2}}.
	\end{aligned}
\end{equation}
Here, $\beta$ is the anisotropy parameter, and the velocity dispersion ellipsoid is assumed to be spherically aligned, giving $\moment{v_r v_\theta} = 0$.

The line-of-sight second moment $\moment{v_{\rm los}^2}$ is the integral given by
\begin{equation}
	S \moment{v_{\rm los}^2} (x, y) = \int_{-\infty}^{\infty} \rmd z\ \zeta \moment{v_z^2}.
\end{equation}
where $S(x, y)$ is the surface density of the dynamical tracer. Given that there is no evidence of significant ordered rotation and the only significantly nonzero velocities are likely due to systematic errors (see \figref{fig:kinematic_maps}), we assume $\moment{v_{\rm los}}=0$ and define $\moment{v_{\rm los}^2}= \sigma_{\rm los}^2$. %to avoid introducing unnecessary noise in the second moments. 
The observed line-of-sight velocity dispersion is given a luminosity-weighted integral as
\begin{equation} \label{eq:observed_dispersion}
	\left[\sigma^2_{\rm los}\right]_{\rm obs} = \left[\moment{v_{\rm los}^2}\right]_{\rm obs} = \frac{\int_{\rm ap} \rmd x\rmd y\ I \moment{v_{\rm los}^2} \otimes {\rm PSF} }{ \int_{\rm ap} \rmd x \rmd y \  I \otimes {\rm PSF}},
\end{equation}
where the symbol ``$\otimes\ {\rm PSF}$'' denotes a convolution with the PSF. In the equation above, we have chosen the surface brightness profile $I(x,y)$ as a substitute for the surface density $S(x, y)$ of the dynamical tracer since the constant factor between surface brightness and surface number density cancels out in this expression.

We use the dynamical modeling software \textsc{jampy}\footnote{\url{https://pypi.org/project/jampy/}} to compute the observed velocity dispersion by solving the Jeans equation from \equref{eq:jeans} for a given 3D potential $\Phi(r)$ and anisotropy profile $\beta(r)$. Specifically, we use the \texttt{jam\_axi\_proj()} routine with the keyword \texttt{align=`sph'}. See \citet{Cappellari08, Cappellari20} for a detailed formalism in computing \equref{eq:observed_dispersion} by \textsc{jampy}.

\subsection{Cosmological inference from combining dynamical and lensing observables}

\ajsthr{We parametrize the 3D potential $\Phi(r)$ using the lens model parameters $\xi_{\rm mass}$ and the internal MST parameter $\lambda_{\rm int}$ to conveniently use the lens model posterior from \citet{Suyu13} as a mass model prior in the dynamical modeling. Thus from \equref{eq:complete_mst}, the surface mass density for our dynamical model is given by}
\begin{equation} \label{eq:full_surface_density_profile}
	\Sigma (\btheta) = \Sigma_{\rm cr} (1 - \kappa_{\rm ext}) \left[ \lambda_{\rm int}\ \kappa_{\rm model}(\btheta) + (1 - \lambda_{\rm int}) \kappa_{\rm s}(\theta) \right].
\end{equation}
\ajsthr{We include $\Ddt$ and $\Dd$ as free parameters in our model, which give the critical density $\Sigma_{\rm cr}$ as}
\begin{equation}
	\Sigma_{\rm cr} = \frac{c^2}{4 \uppi G} \frac{\Ddt}{(1 + z_{\rm d}) \Dd^2} = \frac{c^2}{4 \uppi G} \frac{\Ddt^{\rm model}}{(1 + z_{\rm d}) (1 - \kappa_{\rm ext}) \lambda_{\rm int} \Dd^2},
\end{equation}
\ajsthr{where $\Ddt^{\rm model}$ is the time-delay distance predicted by the lens mass model $\kappa_{\rm model}(\btheta)$ for the time delays observed by \citet{Tewes13}.}

We approximate the surface mass density $\Sigma(\btheta)$ with an MGE \citep{Emsellem94, Cappellari02, Shajib19b} using the software program \textsc{mgefit}\footnote{\url{https://pypi.org/project/mgefit/}}. \textsc{jampy} deprojects the MGE components into an oblate or prolate spheroid with an inclination angle $i$ \citep{Cappellari02}. \ajsthr{The deprojected 3D mass density provides the 3D potential $\Phi$ for the kinematic computation. We also take the MGE of the surface brightness $I(x, y)$ for deprojection to 3D with the inclination angle $i$ for the kinematic computation by \textsc{jampy}.}

\ajsthr{The combination of lens imaging observables and the stellar kinematics is sensitive to $\lambda_{\rm int} (1 - \kappa_{\rm ext}) \Ds/\Dds$ \citep{Birrer16, Chen21}. We apply a prior on $\kappa_{\rm ext}$ using the estimated $\kappa_{\rm ext}$ distribution from \citet{Suyu14} to help break the degeneracy in distributing the total MSD into external and internal components.}

\subsection{Bayesian framework} \label{sec:bayes}

According to Bayes' theorem, the posterior of the model parameters $\Xi = \{\xi_{\rm mass}, \xi_{\rm light}, \Ddt^{\rm model}, i, \kappa_{\rm ext}, \lambda_{\rm int}, \Dd, \beta \}$ as
\begin{equation}
	p (\Xi \mid \mathcal{D} ) \propto p ( \mathcal{D} \mid \Xi)\ p(\Xi),
\end{equation}
where $p(\mathcal{D} \mid \Xi)$ is the likelihood given data $\mathcal{D}$ and $p(\Xi)$ is the prior. In this study, the data $\mathcal{D}$ is the measured velocity dispersions in Voronoi bins (\figref{fig:kinematic_maps}). The observational information from the published time delays, lens models using HST imaging, and the line-of-sight effects \citep{Tewes13, Suyu13, Suyu14} is incorporated by adopting those previous posteriors as the prior on our model parameters. The likelihood of the observed velocity dispersion vector $\bm{\sigma}_{\rm los} \equiv [\sigma_{1}, \dots, \sigma_{N_{\rm bin}} ]$, with $N_{\rm bin}$ being the number of Voronoi bins, is given by
\begin{equation}
	%\begin{aligned}
		\mathcal{L} (\bm{\sigma}_{\rm los} \mid \Xi) \propto \exp \left[ - \frac{1}{2} \bm{\sigma}_{\rm los}^{\rm T} \Sigma^{-1} \bm{\sigma}_{\rm los} \right],
	%\end{aligned}
\end{equation}
where $\Sigma$ is the variance-covariance matrix. Specific priors used in this Bayesian framework are given in \secref{sec:dynamic_models}. We obtain the posterior probability distribution function (PDF) of the model parameters using the Markov-chain Monte Carlo (MCMC) method using the affine-invariant ensemble sampler \textsc{emcee} \citep{Goodman10, Foreman-Mackey13}. We ensure the MCMC chains' convergence by running the chains for $\gtrsim$20 times the autocorrelation length after the chains have stabilized \citep{Foreman-Mackey13}.

\section{Dynamical models} \label{sec:dynamic_models}

We first describe our baseline dynamical model in \secref{sec:baseline_dynamics} and then perform various checks on systematics in \secref{sec:systematic_checks}.

\subsection{Baseline dynamical model} \label{sec:baseline_dynamics}

This subsection describes the baseline settings in our dynamical model, namely the specific parametrization of the mass model (\secref{sec:dynamical_pl_model}), the dynamical tracer profile (\secref{sec:tracer_profile}), the probability of oblate or prolate axisymmetry (\secref{sec:oblate_prolate}), the inclination angle (\secref{sec:inclination}), and the choice of anisotropy profile (\secref{sec:aniostropy_choice}).

\subsubsection{Parametrization of the mass model} \label{sec:dynamical_pl_model}

We adopt the power-law mass model as our baseline model. In this model, the mass profile is defined with Einstein radius $\theta_{\rm E}$, logarithmic slope $\gamma$, projected axis ratio $q_{\rm m}$, and position angle $\varphi_{\rm mass}$. The convergence profile $\kappa_{\rm model}$ in \equref{eq:full_surface_density_profile} for the power-law model is given by
\begin{equation}
	\kappa_{\rm model}^{\rm pl} (\theta_1, \theta_2) = \frac{3 - \gamma}{2}\left(  \frac{\theta_{\rm E}}{\sqrt{q_{\rm m} \theta_1^2 + \theta_2^2 /q_{\rm m}} } \right)^{\gamma - 1}
\end{equation}
Here, the coordinates $(\theta_1,\ \theta_2)$ are rotated by $\varphi_{\rm mass}$ from the (RA, Dec) coordinate system. We adopt the lens model posterior from \citet{Suyu13} as a prior in our dynamical model. For simplicity, we set the position angle $\varphi_{\rm mass}$ the same as the observed position angle of light $\varphi_{\rm light}$. \ajsthr{We use the estimated $\kappa_{\rm ext}$ distribution for the power-law model as the prior \citep[see Figure 3 of][]{Suyu14}.}

\ajsthr{We set $\theta_{\rm s} = 12\arcsec$ ($\simeq 7.5 \theta_{\rm E}$) in the approximate mass-sheet $\kappa_{\rm s}$ (Equation \ref{eq:variable_mass_sheet}) so that the imaging constraints alone cannot differentiate the power-law mass profile and its approximate MST from \equref{eq:approximate_mst}. We obtain this lower limit by running the \textsc{jupyter} notebook that produces Figure 3 of \citep{Birrer20}.\footnote{\url{https://github.com/TDCOSMO/hierarchy_analysis_2020_public/blob/6c293af582c398a5c9de60a51cb0c44432a3c598/MST_impact/MST_pl_cored.ipynb}} However, we adjusted the fiducial lens model parameters in the notebook to match with those for \lensname. We take a uniform prior for the internal MST parameter $\lambda_{\rm int} \sim \mathcal{U}(0.5, 1.13)$. The upper limit of 1.13 is set by the requirement that the transformed mass profile under the approximate MST must be monotonic so that the MGE can approximate the transformed profile sufficiently well \citep{Shajib19}. Previous studies also found similar or more restrictive upper limits for $\lambda_{\rm int}$ to satisfy the physical requirement of non-negative density \citep{Birrer20, Yildirim21}.}

\ajsfou{The appropriate number of MGE components for the mass or light profile is automatically chosen by \textsc{jampy} with a maximum of 20 components.} \ajsfiv{We check that the MGE approximates the input mass or light profile very well (with a maximum 1\% deviation at $< 10\arcsec$ and maximum 10\% deviation between 10$\arcsec$--50$\arcsec$). These deviations from the density profile have an oscillatory pattern due to the MGE approximation's nature, except near the end of the fitted ranges. Thus the deviation in the integrated mass profile often averages out in the line-of-sight integration up to a very large radius. We perform the MGE fitting up to 100$\arcsec$. Thus, the large mismatch between the MGE approximation and the original profile occurs largely outside the integration limit $\sim70\arcsec$. The chosen number of maximum Gaussian components is not a dominant source of numerical error. Setting this maximum number to a very high value, such as 100, shifts the computed velocity dispersion by only $<0.5$\% within the observed region, which is insignificant compared to the 1\% numerical stability targeted by \textsc{jampy}.}

\subsubsection{Dynamical tracer profile} \label{sec:tracer_profile}
\begin{figure*}
	\includegraphics[width=\textwidth]{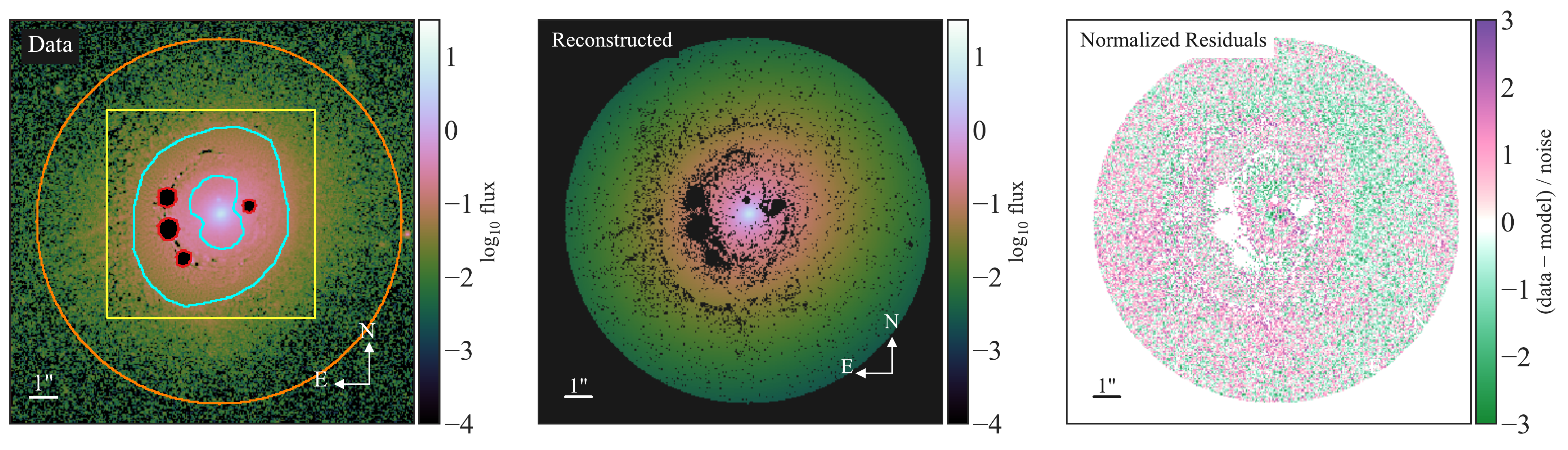}
	\caption{\label{fig:light_fitting}
	Fit of the lens galaxy's surface brightness profile. \textbf{Left:} the \ajsthr{HST/ACS imaging in the F814W filter} of the lens system \lensname\ with the quasar images and the lensed arcs subtracted using the prediction from the best-fit lens model from \citet{Suyu13}, thus leaving only the lens galaxy's light to be fitted. The orange circle shows the large circular region considered for fitting in our analysis, and the yellow square shows the smaller cutout used for lens modeling by \citet{Suyu13}. The cyan annulus contains the region where pixels were fitted to reconstruct the source by \citet{Suyu13}. Thus the lensed arcs from the quasar host galaxy were subtracted only within this annulus. The red contours mark quasar image positions with significant residuals due to saturated pixels, which we mask. \textbf{Middle:} The fitted light profile with a double S\'ersic model. The black pixels correspond to masked pixels. The additional masked pixels within the orange circle not described above are randomly selected through an iterative process that performs outlier rejection while preserving the Gaussian tail (see \secref{sec:tracer_profile} for details). \textbf{Right:} Normalized residual of the best-fit model.
	}
\end{figure*}

We update the light profile fitting for the lens galaxy from \citet{Suyu13} using a larger HST image cutout than that therein, which did not contain the full extent of the lens galaxy's light profile (see \figref{fig:light_fitting}). The lensed arcs and quasar images are first subtracted from the cutout using the prediction of the best-fit lens model from \citet{Suyu13}. \ajstwo{We use the software package \textsc{lenstronomy}\footnote{\url{https://github.com/lenstronomy/lenstronomy}} to fit the residual light distribution attributed to the lens galaxy \citep{Birrer18, Birrer21b}.} Following \citet{Suyu13}, we use the double S\'ersic model to fit the light profile, which is a superposition of two concentric S\'ersic profiles. The S\'ersic profile is defined as
\begin{equation}
	I(\theta_1, \theta_2) = I_{\rm 0} \exp \left[ -b_n \left( \frac{\sqrt{q_{\rm l} \theta_1^2 + \theta_2^2/q_{\rm l}}}{\theta_{\rm eff}}\right)^{1/n_{\rm s}} + b_n \right],
\end{equation}
where $I_0$ the amplitude, $q_{\rm l}$ is the axis ratio, $\theta_{\rm eff}$ is the effective radius, $n_s$ is the S\'ersic index, and $b_n = 1.999n - 0.327$ is a normalizing factor so that $\theta_{\rm eff}$ becomes the half-light radius \citep{Sersic68}. The coordinates $(\theta_1,\ \theta_2)$ are rotated by $\varphi_{\rm light}$ from the (Ra, Dec) coordinate system. 

We first mask circular regions at the quasar image positions due to slightly saturated pixels producing significant residuals in the subtracted cutout (see \figref{fig:light_fitting}). We then iteratively mask the other pixels with significant residuals above statistical expectations to effectively perform an outlier rejection while preserving the shape of a Gaussian tail. For each iteration of this process, we take a discrepancy threshold, which we decrease from 5$\sigma$ to 2$\sigma$ with step size 0.5$\sigma$ across these iterations. We then randomly mask a subset of the pixels with residuals more than the discrepancy level at the given iteration such that the number of remaining pixels with such high residuals is statistically expected. The final masked area after the iterations is illustrated in \figref{fig:light_fitting}. We tabulate the best-fit light model parameters in \tabref{tab:light_parameters} and compare them with those from \citet{Suyu13}. The circularized half-light radius for our best-fit model is $\theta_{\rm eff} = 1\farcs91$, which is slightly larger than the value $\theta_{\rm eff}=1\farcs85$ from \citet{Suyu13} \ajsthr{based on the same imaging data but from a smaller cutout (illustrated in \ref{fig:light_fitting})}. We then take the MGE of the fitted double S\'ersic profile as the light distribution $I(x, y)$ in our dynamical modeling. We propagate the uncertainties and covariances from the light profile fitting into the dynamical modeling. To do that, we sample from the multivariate normal distribution corresponding to all the light model parameters for each call of the likelihood function within the MCMC process and then take the MGE of the light profile given the sampled parameters.

\renewcommand*\arraystretch{1.15}
\begin{table}
\caption{ \label{tab:light_parameters}Values of the light model parameters for the double S\'ersic model in our fitting of a large cutout and those from \citet{Suyu13}. The position angle $\varphi_{\rm light}$ is defined as East of North.}
\begin{minipage}{\linewidth}
\begin{tabular}{lcc}
\hline
Parameter &
This analysis &
\citet{Suyu13} \\
\hline
\multicolumn{3}{c}{S\'ersic profile 1} \\
\hline
$I_0$ (e$^{-1}$ s$^{-1}$ pixel$^{-1}$) & 
32.8 \pml 0.1 &
36.4 \pml 0.4
\\
$\theta_{\rm eff}$ (\arcsec) &
2.437 \pml 0.005 &
2.49 \pml 0.01
\\
$n_{\rm s}$ &
1.10 \pml 0.01 &
0.93 \pml 0.03
\\
$q_{\rm l}$ &
0.865 \pml 0.001 &
0.878 \pml 0.004
\\
\hline
\multicolumn{3}{c}{S\'ersic profile 2} \\
\hline
$I_0$ (e$^{-1}$ s$^{-1}$ pixel$^{-1}$) & 
441 \pml 7 &
356 \pml 12
\\
$\theta_{\rm eff}$ (\arcsec) &
0.300 \pml 0.003 &
0.362 \pml 0.009
\\
$n_{\rm s}$ &
1.60 \pml 0.02 &
1.59 \pml 0.03
\\
$q_{\rm l}$ &
0.847 \pml 0.002 &
0.849 \pml 0.004
\\
$\varphi_{\rm light}$ ($^\circ$) &
120.5 \pml 0.3 &
121.6 \pml 0.5
\\
\hline
\end{tabular}
\end{minipage}
\end{table}
\renewcommand*\arraystretch{1.0}

\subsubsection{Oblate or prolate shape of the axisymmetry} \label{sec:oblate_prolate}

\begin{figure}
    \includegraphics[width=\columnwidth]{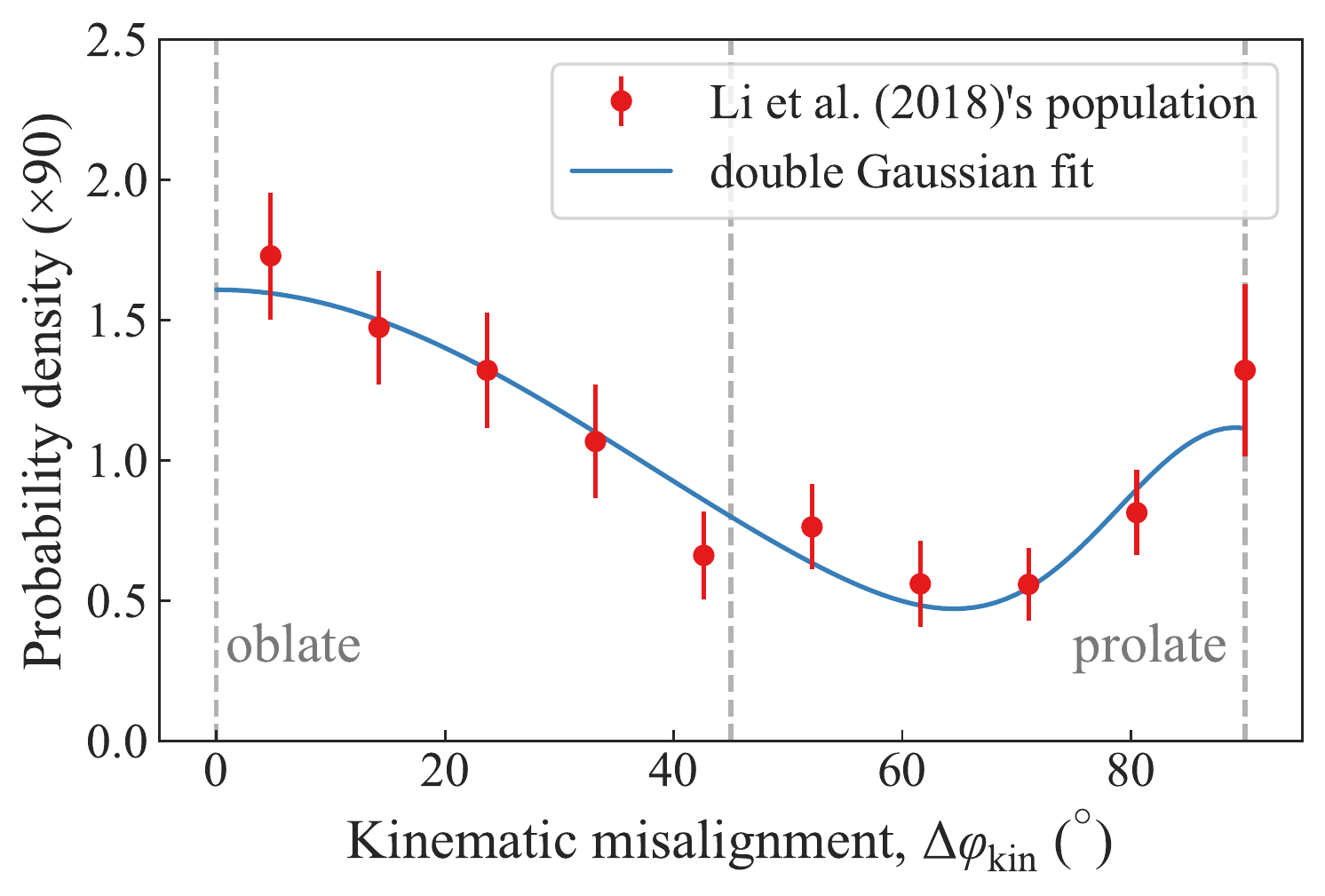}
	\caption{\label{fig:oblateness}
	Population prior on kinematic misalignment angle $\Delta \varphi_{\rm kin} \equiv \left| \varphi_{\rm kin} - \varphi_{\rm light} \right|$ for a sample of slow rotator elliptical galaxies from the SDSS's MaNGA dataset \citep{Li18}. Here, $\Delta \varphi_{\rm kin} = 0^{\circ}$ corresponds to a purely oblate shape, and $\Delta \varphi_{\rm kin} = 90^{\circ}$ corresponds to a purely prolate shape. The vertical dashed grey lines mark $\Delta \varphi_{\rm kin} = 0^\circ$, $45^{\circ}$, and $90^\circ$. The red points with error bars show the measurements from \citet{Li18}. We fit this distribution with a double Gaussian model (blue line) with the means fixed to  $\Delta \varphi_{\rm kin} = 0^{\circ}$ and $\Delta \varphi_{\rm kin} = 90^{\circ}$. We take $\Delta \varphi_{\rm kin} < 45^\circ$ as the oblate case and $\Delta \varphi_{\rm kin} > 45^{\circ}$ as the prolate case. Integrating the double Gaussian model from $0^{\circ}$ to $45^{\circ}$ gives the prior probability of oblateness $p({\rm oblate})_{\rm pop} \simeq 0.65$.}
\end{figure}

The oblateness, prolateness, or triaxiality of a slow rotator galaxy can, in principle, be constrained from the kinematic misalignment angle $\Delta \varphi_{\rm kin} \equiv \left| \varphi_{\rm kin} - \varphi_{\rm light} \right|$. However, we do not detect any significant rotational pattern in the $v_{\rm mean}$ map (\figref{fig:kinematic_maps}). Thus, the uncertainty for the constrained kinematic major axes is too large to be meaningful, and we cannot directly constrain this galaxy's oblateness from the data. Instead, we obtain the probability of oblateness from a population prior based on 189 slow rotator elliptical galaxies that are in the Sloan Digital Sky Survey's (SDSS's) Mapping Nearby Galaxies at APO (MaNGA) sample \citep{Abolfathi18, Graham18}. We take the distribution of $\Delta \varphi_{\rm kin}$ for this sample of slow rotators \citep{Li18}, where $\Delta \varphi_{\rm kin} = 0^{\circ}$ corresponds to a purely oblate shape, and $\Delta \varphi_{\rm kin} = 90^{\circ}$ corresponds to a purely prolate shape. \citet{Li18} find two distinct peaks in the distribution at $\Delta \varphi_{\rm kin} = 0^{\circ}$ and $\Delta \varphi_{\rm kin} = 90^{\circ}$ (see \figref{fig:oblateness}). We, therefore, fit the data points with a double Gaussian profile with the means set at $\Delta \varphi_{\rm kin} = 0^{\circ}$ and $\Delta \varphi_{\rm kin} = 90^{\circ}$ (see the fit in \figref{fig:oblateness}). Although the slow rotators with $0^{\circ} < \Delta \varphi_{\rm kin} < 90^{\circ}$ have triaxial shapes, we choose only oblate or prolate axisymmetric shapes in our dynamical modeling for computational simplicity. Therefore, we take $\Delta \varphi_{\rm kin} < 45^\circ$ as the oblate case and $\Delta \varphi_{\rm kin} > 45^\circ$ as the prolate case. We obtain the prior probability $p({\rm oblate})_{\rm pop}$ of the galaxy being oblate as
\begin{equation}
	\ajstwo{p({\rm oblate})_{\rm pop} = \int_{0^\circ}^{45^\circ} {\rmd} (\Delta \varphi_{\rm kin})\ p(\Delta \varphi_{\rm kin})_{\rm pop} \simeq 0.65,}
\end{equation}
\ajstwo{and thus $p({\rm prolate})_{\rm pop} = 1 - p({\rm oblate})_{\rm pop} \simeq 0.35$.}

The \textsc{jampy} software package, by default, adopts the oblate case for deprojection. We implement the prolate case in \textsc{jampy} by setting $q_{\rm prolate} = 1 / q > 1$ and switching the $x$ and $y$ axes in the input coordinate system. \ajstwo{Due to the switching of $x$ and $y$ axes, $\sigma$ parameters of the MGEs for mass and light models need to be scaled as $\sigma_{\rm prolate} = q\sigma$.}

\subsubsection{Inclination} \label{sec:inclination}
\begin{figure}
	\includegraphics[width=\columnwidth]{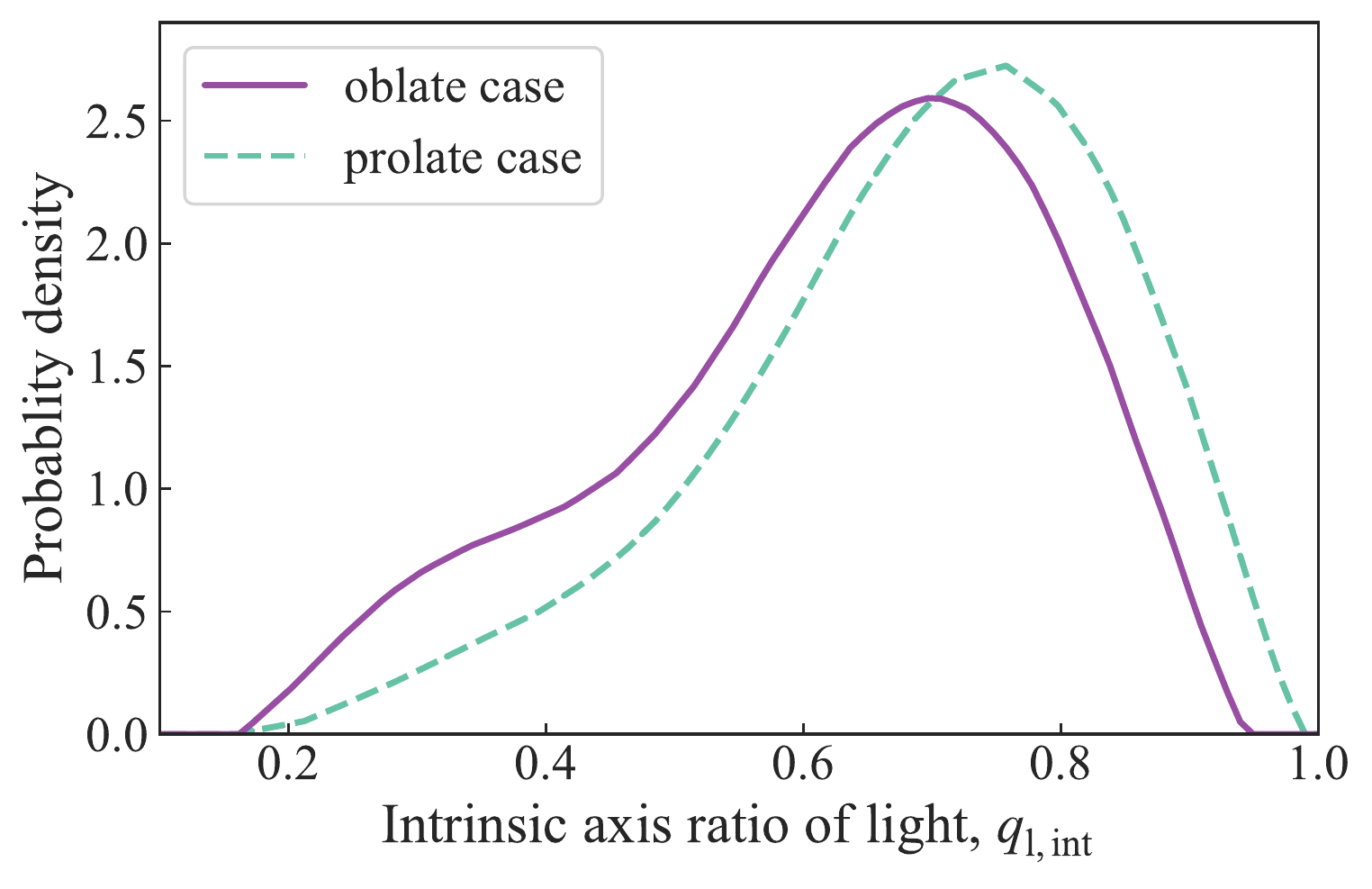}
	\caption{\label{fig:q_intrinsic}
	Prior on the intrinsic axis ratio $q_{\rm l, int}$ of light for oblate (solid line) and prolate (dashed line) cases from \citet{Chang13}. The priors correspond to massive elliptical galaxies from the SDSS survey at $0.04 < z < 0.08$ with $10.8 < \log_{10}(M_{\star}/M_{\odot}) < 11.5$.
	}
\end{figure}

The observed axis ratio \ajsthr{of light} $q_{\rm l, obs} = 0.850 \pm 0.002$ relates to $q_{\rm l, int}$ through the inclination angle $i$ as 
\begin{equation}
	q_{\rm l, obs}^2 =	q_{\rm l, int}^2 \sin^2 i + \cos^2 i.
\end{equation}
We impose a prior on the intrinsic axis ratio $q_{\rm l, int}$ from a sample of massive elliptical galaxies in the SDSS with stellar mass $10.8 < \log_{10}(M_{\star}/M_{\odot}) < 11.5$ at $0.04 < z < 0.08$ \citep{Chang13}. \ajsthr{The distribution of $q_{\rm l, int}$ by \citet{Chang13} is different for oblate and prolate assumptions. Therefore, we adopt \ajsthr{the specific prior} corresponding to the oblate or the prolate case (see \figref{fig:q_intrinsic}).}

\subsubsection{Anisotropy profile} \label{sec:aniostropy_choice}

\begin{figure*}
	\includegraphics[width=\textwidth]{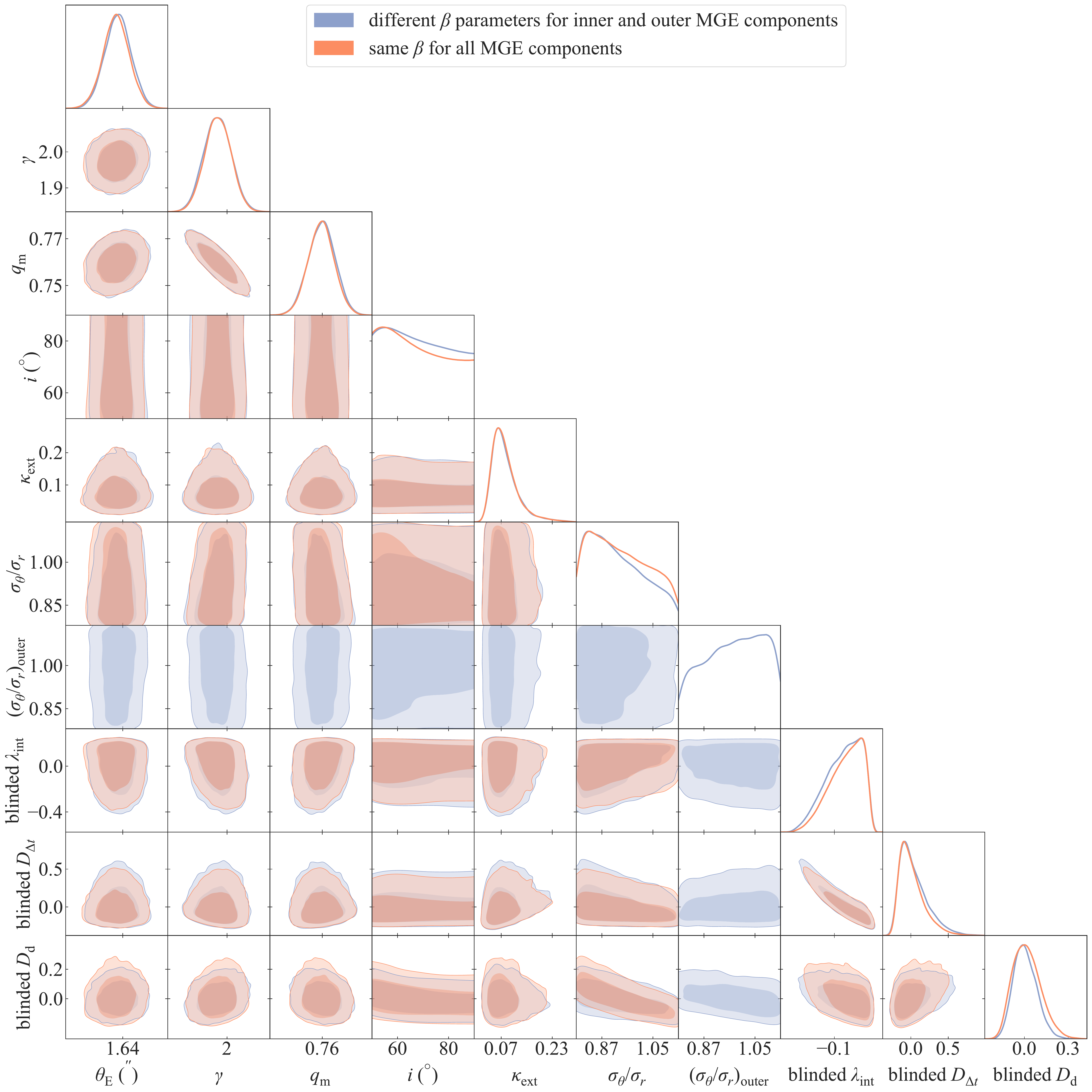}
	\caption{\label{fig:axisymmetric_jam_corner_powerlaw}
	Constraints from axisymmetric JAM modeling on the power-law mass model parameters ($\theta_{\rm E}$, $\gamma$, and $q_{\rm m}$), internal MST parameter $\lambda_{\rm int}$, external convergence $\kappa_{\rm ext}$, anisotropy profile parameter(s), and the cosmological distances $\Ddt$ and $\Dd$. assuming two anisotropy parametrizations: (i) one single constant $\beta \equiv 1 - (\sigma_{\theta}/\sigma_r)^2$ for all light MGE components (orange contours), and (ii) one free $(\sigma_{\theta}/\sigma_r)_{\rm inner} \equiv (\sigma_{\theta}/\sigma_r)$ for light MGE components with $\sigma < r_{\rm break} =  \theta_{\rm eff} = 1\farcs91$ and another free $(\sigma_{\theta}/\sigma_r)_{\rm outer}$ for light MGE components with $\sigma > r_{\rm break}$(blue contours). The blinded parameters are blinded as $p_{\rm blinded} \equiv p/\langle p \rangle - 1$ so that the distributions only reveal fractional uncertainties. The darker and lighter shaded regions in the 2D plots trace 68\% and 95\% credible regions, respectively. The mass model parameters Einstein radius $\theta_{\rm E}$, power-law slope $\gamma$, axis ratio $q$, and position angle PA are additionally constrained through a prior from the imaging data from \citet{Suyu13}. The two anisotropy parametrizations provide equally good fits to the kinematics data. However, the BIC selects the constant-$\beta$ anisotropy model over the other one with one additional free parameter ($\Delta$BIC value is \deltabicani).
	}
\end{figure*}

\begin{figure*}
	\includegraphics[width=\textwidth]{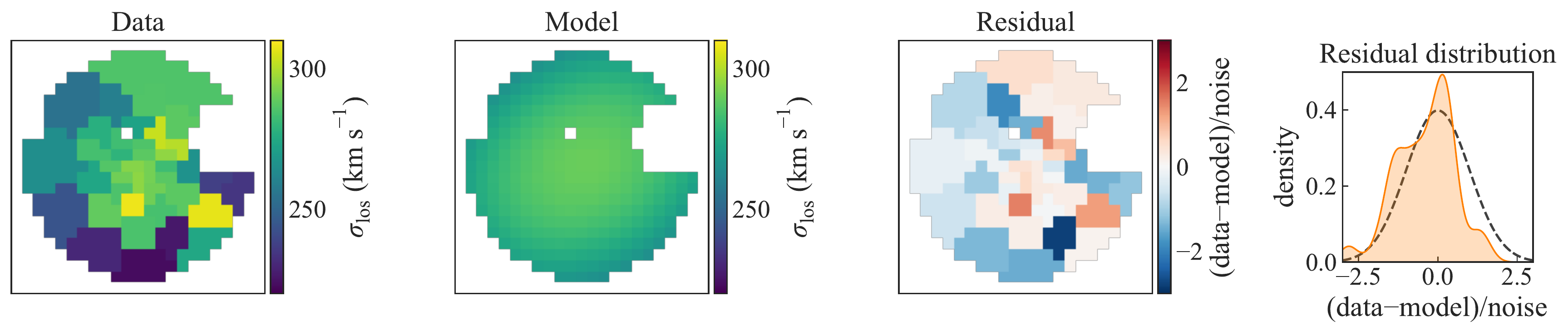}
	\caption{\label{fig:jampy_pl_fit}
	Observed velocity dispersion map in Voronoi bins (first panel), the best-fit dynamical model with a power-law mass model, constant $\beta$ anisotropy profile, and oblate shape (second panel), the normalized residual for the best-fit dynamical model (third panel), and the distribution of the normalized residual (orange, fourth panel). The reduced $\chi^2$ quantity is ${\chi^2_{\nu}} = 0.83$ with degrees of freedom $\nu = 41$. \ajsfou{The grey dashed line in the fourth panel shows a normal distribution expected for residuals from a perfect model to the data with Gaussian noise. The residual distribution for 41 points is similar to this Gaussian distribution.}
	}
\end{figure*}

\begin{figure}
    \centering
    \includegraphics[width=\columnwidth]{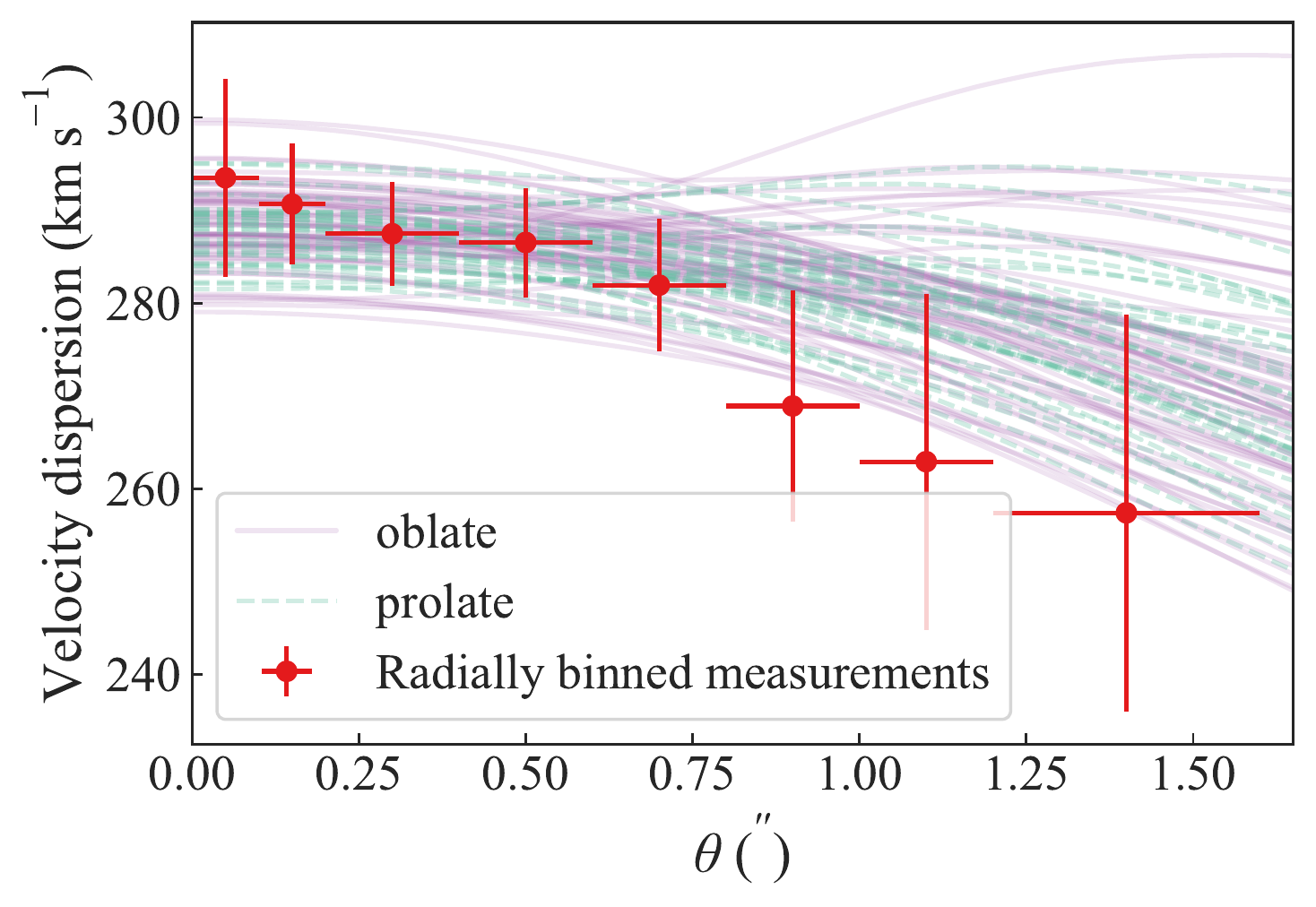}
    \caption{\ajsthr{Radial profile of the line-of-sight velocity dispersion. The red points are radially binned values from the 2D maps, with the horizontal error bars illustrating the widths of the annuli. The lines show the radial profiles for random samples from the dynamical model posterior. The radial profile of the model is averaged over the major, minor, and intermediate axes. The solid purple lines correspond to 65 random samples for the oblate case, and the dashed green lines correspond to 35 random samples for the prolate case. Note that the model was fit to the 2D kinematics data. However, we illustrate the 1D radial profile only for visualization.}
    }
    \label{fig:vel_dis_radial_profile}
\end{figure}

We investigate two choices to parametrize the anisotropy profile. The first choice is a \ajsthr{single spatially constant $\beta = 1 - \sigma_{\theta}^2 / \sigma_{r}^2$ value for all the light MGE components}. \ajsthr{Numerically, we sample $\sigma_{\theta} / \sigma_{r}$ with a uniform prior $(\sigma_{\theta} / \sigma_{r}) \sim \mathcal{U}(0.78, 1.14)$. This range of $\sigma_{\theta} / \sigma_{r}$ allows $-0.31 < \beta < 0.38$. We adopt this range using the $\beta$ values of eight slow rotator galaxies measured by \citet[][see Figure 2]{Cappellari07}. These measurements of $\beta$ by \citet{Cappellari07} are from Schwarzschild modeling of data with one of the highest $S/N$ values in the literature, allowing to constrain the Gauss--Hermite moments up to order six. Applying the student's $t$-distribution on the sample mean of this small sample, we find the 95\% confidence interval of the population mean for $\beta$ to be [-0.10, 0.17] and the standard deviation to be 0.16. These values infer that 95\% of the population is contained within $\beta \in [-0.31, 0.38]$, which we take as the boundaries of our prior range.} The second choice of the anisotropy profile has two free parameters: the \ajsthr{inner light MGE components} with $\sigma < r_{\rm break} = \theta_{\rm eff} = 1\farcs91$ are assigned one value for $(\sigma_{\theta} / \sigma_{r})_{\rm inner}$ and the outer light MGE components with $\sigma \geq r_{\rm break}$ are assigned another independent value of $(\sigma_{\theta} / \sigma_{r})_{\rm outer}$. \ajsthr{Thus, this parametrization with two free parameters allows radial variability in the anisotropy profile. Both the inner and outer ratios have uncorrelated uniform priors $(\sigma_{\theta} / \sigma_{r}) \sim \mathcal{U}(0.78, 1.14)$.} For these two choices of parametrization, we compute the Bayesian information criterion (BIC) given by
\begin{equation}
	{\rm BIC} \equiv k \log (N_{\rm bin}) - 2 \log \hat{\mathcal{L}}, 
\end{equation}
where $k$ is the number of free model parameters, $N_{\rm bin}$ is the number of data points, and $\hat{\mathcal{L}}$ is the maximum likelihood. \ajsthr{We approximate $\hat{\mathcal{L}}$ from the highest likelihood value sampled in the MCMC chain.} The single-parameter $\beta$ model provides the lowest BIC value excluding the two-parameter $\beta$ model with $\Delta{\rm BIC} \approx 3.7$ \citep[i.e., positively excluded;][]{Raftery95}. \ajsfiv{We check that the difference between the highest and the second highest likelihood values among the MCMC samples is $\ll \Delta$BIC, thus this $\Delta$BIC value is robust against our approximation of $\hat{\mathcal{L}}$ from the highest likelihood value in the sampled chain.} The non-detection of varying anisotropy in our data is consistent with that observed in nearby elliptical galaxies, as even high-$S/N$ SAURON data for a large sample of galaxies are accurately described by JAM models with constant anisotropy, as used here, within the noise of the kinematics \citep[e.g.,][]{Cappellari13}. We compare the posterior distributions of the model parameters for the two anisotropy models in \figref{fig:axisymmetric_jam_corner_powerlaw}. An example of a best-fit kinematic model and the corresponding residual with the single-parameter $\beta$ model and oblate axisymmetry is illustrated in \figref{fig:jampy_pl_fit}. \ajsfou{The reduced $\chi_{\nu}^2$ value is 0.83 with $\nu=41$ degrees of freedom. The distribution of residuals is similar to a normal distribution expected from a perfect model for data with Gaussian noise, illustrating that our model is appropriate for the data.} \ajsthr{We show the range of velocity dispersion radial profiles sampled by our model in \figref{fig:vel_dis_radial_profile} and compare it with the radially averaged measurements of the velocity dispersion. This illustration shows that our model reproduces the uncertainty range of the measurement.}

\subsection{Checking potential systematics due to modeling choices} \label{sec:systematic_checks}

In this section, we perform several checks on potential systematics for different choices in the dynamical model setup.

\subsubsection{Comparison between power-law and composite mass models}

In addition to the power-law mass model, \citet{Suyu14} also adopted a composite mass model individually describing the lens galaxy's dark matter and baryonic components. The dark matter distribution was modeled with an elliptical NFW profile in the potential. The parameters in this profile are the normalization of the NFW component $\kappa_{\rm s}$, the NFW scale radius $r_{\rm scale}$, and the mass axis ratio $q_{\rm m}$. The baryonic component was modeled with a mass-follow-light profile with a free mass-to-light ratio ($M/L$) parameter. \ajsthr{Thus, this mass model parametrization has one more free parameter than the power-law model. See \citet{Suyu14} for parametric definitions of these profiles.} We implement this composite mass profile as $\kappa_{\rm model}^{\rm comp}$ in \equref{eq:full_surface_density_profile} and adopt the model posterior from \citet{Suyu14} as a prior in our model. \ajsfou{We appropriately convert the ellipticity defined in the potential by \citet{Suyu14} to an ellipticity defined in the convergence in our model.} We take the MGE of this composite surface density model as done for the power-law surface density model. \ajsfou{However, since the dark matter and baryonic components have different ellipticities, we take the MGE of each component separately to preserve the ellipticity information in deprojection. Specifically, We take the MGE of the approximate MST with $\lambda_{\rm int}$ of the dark matter profile and the MGE of an accordingly rescaled baryonic profile, which effectively results in the total mass profile being transformed as the approximate MST with $\lambda_{\rm int}$.}

\begin{figure}
	\includegraphics[width=\columnwidth]{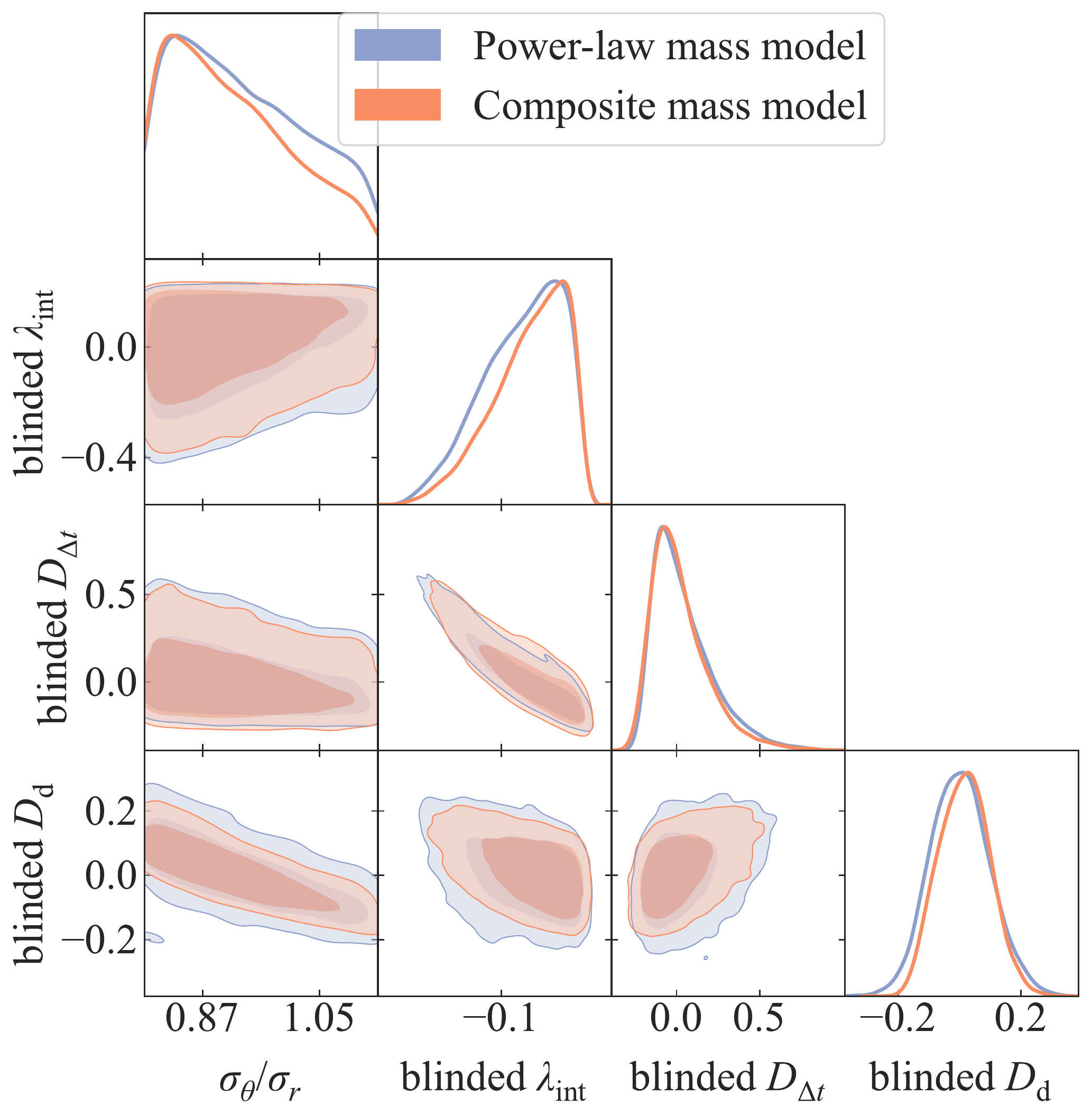}
	\caption{\label{fig:comp_vs_pl_compare}
	Comparison of the constrained $\Dd$ from power-law (blue contours) and composite (orange contours) mass models. The blinded parameters are blinded as $p_{\rm blinded} \equiv p/\langle p \rangle - 1$ so that the distributions only reveal fractional uncertainties. The darker and lighter shaded regions in the 2D plots trace 68\% and 95\% credible regions, respectively.
	}
\end{figure}

This mass model with one more free parameter than the power-law model has a higher BIC score with $\Delta$BIC $=\deltabicpl$. Thus, the BIC excludes the composite model with positive evidence \citep{Raftery95}. The median values of $\Dd$ from the power-law and composite mass models differ by 0.9\% ($0.07\sigma$, \figref{fig:comp_vs_pl_compare}), and the median $\Ddt$ values differ by 1.26\% ($0.06\sigma$). \ajsfou{Therefore, we conclude that our power-law mass model with an additional degree of freedom to scale with the MST robustly describes the observed data.}

\subsubsection{Comparison between prolate and oblate axisymmetry}

\begin{figure}
	\includegraphics[width=\columnwidth]{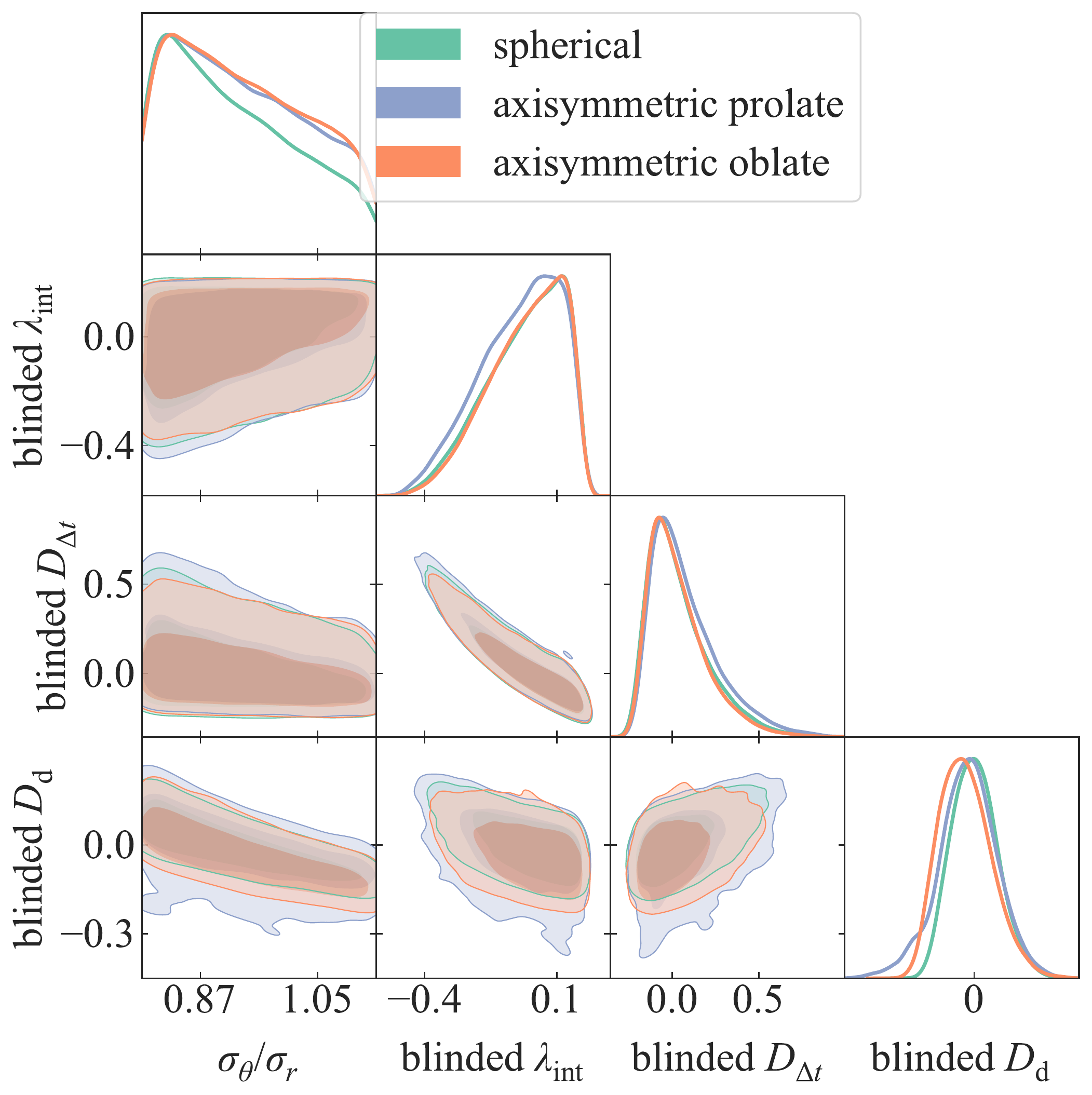}
	\caption{\label{fig:oblateness_vs_prolateness}
	Comparison of the constrained $\Dd$ between oblate (blue) and prolate (blue) cases of the deprojected spheroidal shape in the dynamical model. The blinded parameters are blinded as $p_{\rm blinded} \equiv p/\langle p \rangle - 1$ so that the distributions only reveal fractional uncertainties. The darker and lighter shaded regions in the 2D plots trace 68\% and 95\% credible regions, respectively.
	}
\end{figure}

We compare the inferred $\Dd$ between the purely oblate and purely prolate cases in the deprojected 3D spheroidal shape of the mass and light models (\figref{fig:oblateness_vs_prolateness}). The median $\Dd$ values from these two cases differ by 3.6\% ($0.3\sigma$), and the median $\Ddt$ values differ by 0.94\% ($0.04\sigma$). Our final distance posterior is the combination of oblate and prolate cases, with weights $p({\rm oblate})_{\rm pop} = 0.65$ and $1 - p({\rm oblate})_{\rm pop} = 0.35$, respectively. Thus, this difference between the oblate and prolate cases is marginalized in our final cosmological distance posterior.

We also compare the predictions from axisymmetric and spherical mass models in \figref{fig:oblateness_vs_prolateness}. The median $\Dd$ from the spherical model matches very well with the axisymmetric prolate model, but the median $\Ddt$ differs by 2.0\% ($0.08\sigma$). \ajsfou{The galaxy is only mildly elliptical in projection ($q_{\rm l} \sim 0.85$), and the resulting axisymmetric models are not very flat. For this reason, the relatively small difference between the axisymmetric and spherical models is not surprising.}

\subsubsection{Comparison between Voronoi binning schemes}

\begin{figure}
	\includegraphics[width=\columnwidth]{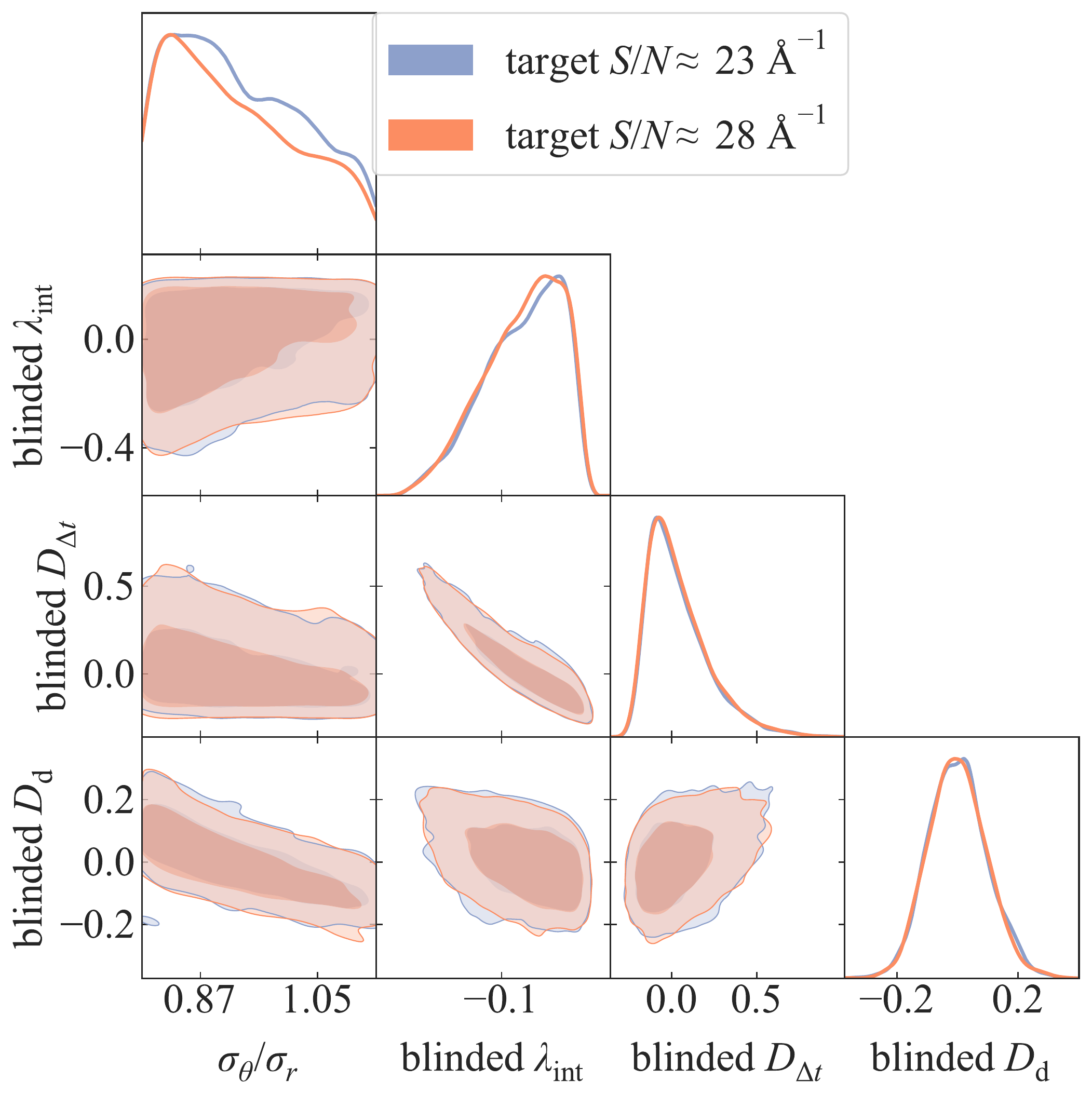}
	\caption{\label{fig:snr_compare}
	Comparison of the constrained $\Dd$ between two choices of the target $S/N$ for each bin in the Voronoi binning scheme. The blinded parameters are blinded as $p_{\rm blinded} \equiv p/\langle p \rangle - 1$ so that the distributions only reveal fractional uncertainties. The darker and lighter shaded regions in the 2D plots trace 68\% and 95\% credible regions, respectively.
	}
\end{figure}

Here, we compare the Voronoi binning schemes with two choices for the target $S/N$ in each bin: \ajsthr{$\approx23$ \AA$^{-1}$ and $\approx28$ \AA$^{-1}$}. The two cases match very well with only a 0.21\% difference ($0.02\sigma$) in the median values of $\Dd$ (\figref{fig:snr_compare}) and 0.28 \% difference (0.01$\sigma$) in the median $\Ddt$ values. As a result, we conclude that our choice of the Voronoi binning scheme is not a significant source of systematic error in our analysis.

\ajsfiv{Based on the systematics tests performed above, we adopt a robust final distance posterior from the model with the power-law parametrization for the mass profile that the approximate internal MST is applied to. We marginalize the oblate and prolate axisymmetrical cases by combining the posteriors from these two choices with weights of 0.65 and 0.35, respectively. In the next section, we present the unblinded values from the distance posterior and infer the value of  \Ho\ from it.}

\section{Cosmological inference} \label{sec:cosmo_inference}

In this section, we infer cosmological parameters from the joint distribution of $\Dd$ and $\Ddt$, accounting for their covariance. The unblinded point estimates of these distances are $\Dd = \Ddval$ Mpc (a 9.6\% measurement) at $z_{\rm d}=0.295$, and $\Ddt = \Ddtval$ Mpc (a 17\% measurement) for $z_{\rm s} = 0.657$.

\begin{figure*}
	\includegraphics[width=\columnwidth]{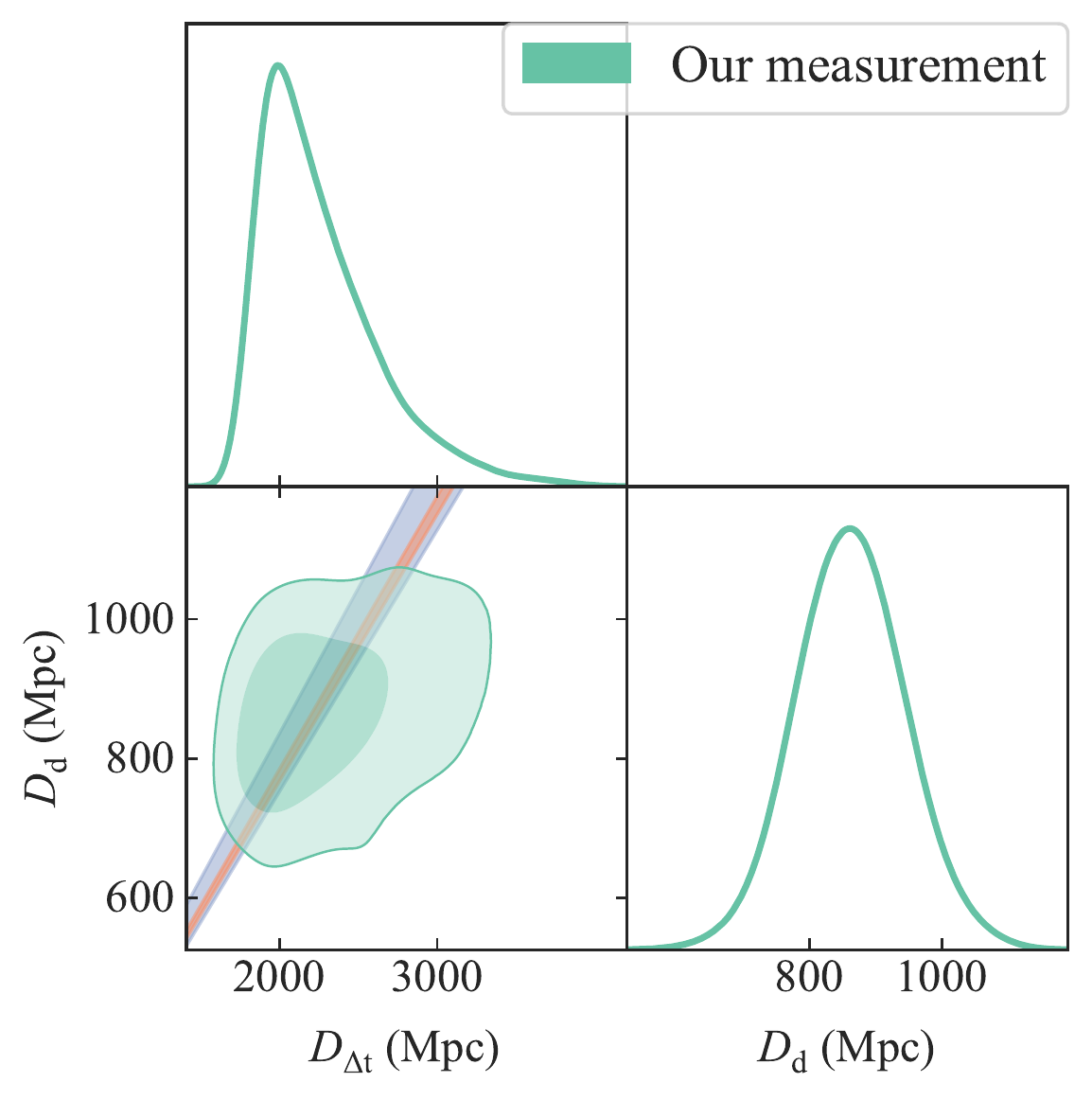}
	\includegraphics[width=\columnwidth]{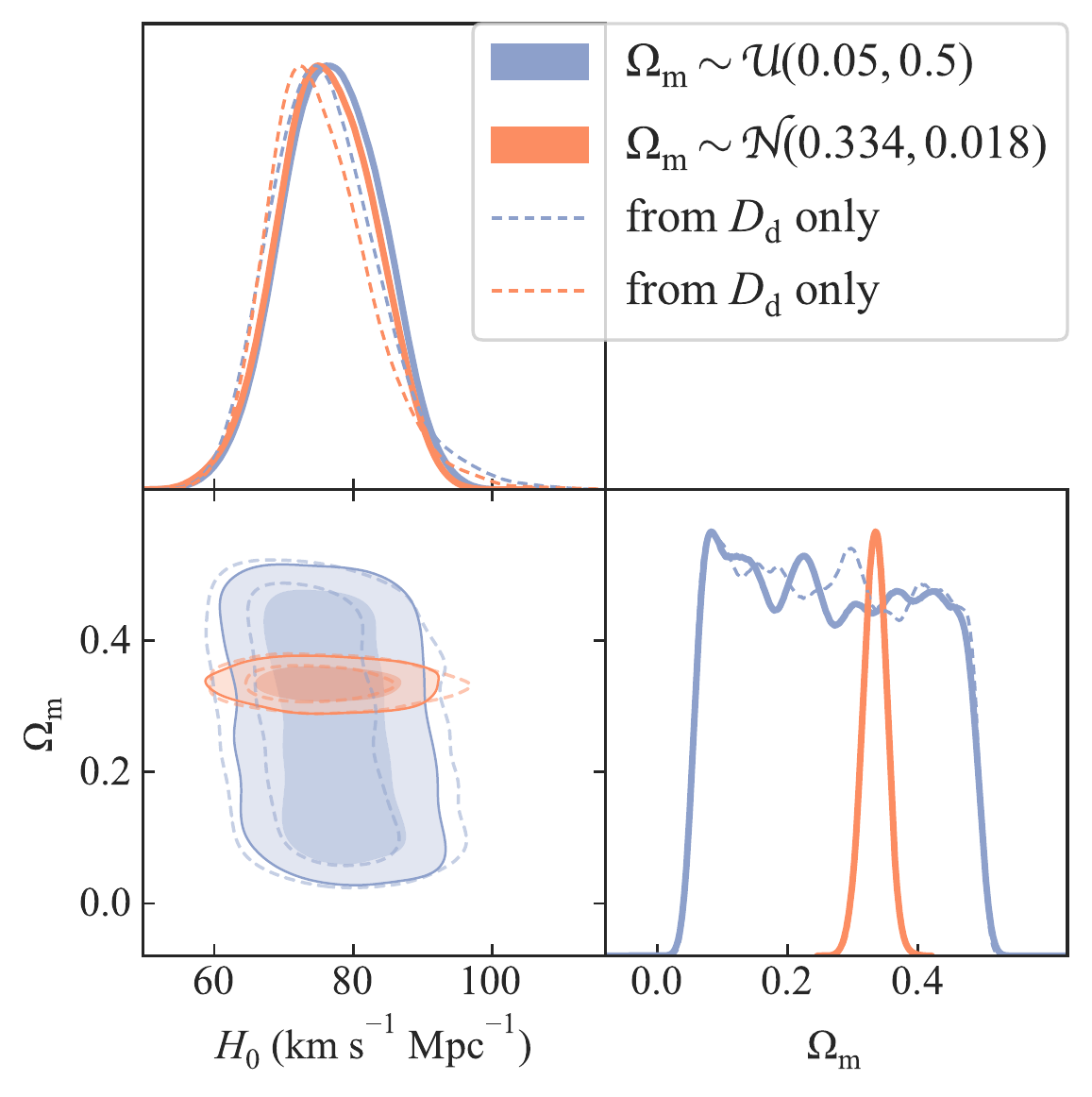}
	\caption{\label{fig:h0}
	\textbf{Left:} Final 2D posterior of the time-delay distance $\Ddt$ and the angular diameter distance $\Dd$ (emerald contour). The darker and lighter shaded regions in the 2D plots trace 64\% and 95\% credible regions, respectively. We infer $\ho$ and $\Omega_{\rm}$ from this distance posterior accounting for the covariance in a flat \lcdm\ cosmology. We take a wide uniform prior on $\ho \sim \mathcal{U}(0, 150)$ \hunit. The blue-shaded region corresponds to a uniform prior $\Omega_{\rm m} \sim \mathcal{U}(0.05,\ 0.5)$ and the orange-shaded region corresponds to a Gaussian prior $\Omega_{\rm m} \sim \mathcal{N} (0.334,\ 0.018)$ from the Pantheon$+$ analysis of type Ia supernovae relative distances \citep{Brout22}. \textbf{Right:} Posterior PDF of $H_0$ and $\Omega_{\rm m}$ in flat \lcdm\ cosmology. We constrain \Ho\ to 9.4\% and 9.1\% precision for the uniform and Pantheon$+$ $\Omega_{\rm m}$-priors, respectively. We show the cosmological parameter posterior from only the 1D $\Dd$ posterior with dashed contours with colors matching the associated $\Omega_{\rm}$ prior. In this case, the \Ho\ precision is 10.3\% and 9.6\% for the uniform and Gaussian priors, respectively. The $\Dd$-only constraint on the \Ho\ is lower by $\sim$1.4\% (0.15$\sigma$) than the constraint from the full 2D posterior, for the uniform $\Omega_{\rm m}$-prior.
	}
\end{figure*}

We infer \Ho\ and $\Omega_{\rm m}$  from our distance posterior for a flat \lcdm\ cosmology (see \figref{fig:h0}, left panel). We leave the exploration of more exotic cosmologies based on our distance posterior for future studies. We approximate the likelihood function $\mathcal{L} (H_0,\Omega_{\rm m} \mid \Dd, \Ddt)$ of the cosmological parameters using a 2D Gaussian kernel density estimate (KDE) from the 2D distance posterior. We adopt two choices of prior for $\Omega_{\rm m}$: one is a uniform prior $\Omega_{\rm m} \sim \mathcal{U}(0.05,\ 0.5)$, and the other is a Gaussian prior $\Omega_{\rm m} \sim \mathcal{N} (0.334,\ 0.018)$ from the Pantheon$+$ analysis of type Ia supernovae relative distances \citep{Brout22}. We infer the posterior joint PDF of \Ho\ and $\Omega_{\rm m}$ by performing MCMC sampling using \textsc{emcee}, given the likelihood function and prior choice.

We infer \Ho\ = $\hval$ \hunit (a 9.4\% measurement) with the uniform $\Omega_{\rm m}$-prior, and \Ho\ = $76.0_{-6.6}^{+7.3}$ \hunit\ (a 9.1\% measurement) with the Pantheon$+$ $\Omega_{\rm m}$-prior (solid contours in the right panel of \figref{fig:h0}). We show the $\Ddt$--$\Dd$ region allowed by our priors in the left panel of Figure \ref{fig:h0}, which also shows the region allowed by our distance posterior that provides information for the cosmological inference. Other cosmological models beyond flat \lcdm\ \citep[e.g.,][]{Bonvin17, Wong20} or combining other cosmological probes in a cosmology-independent manner \citep[e.g.,][]{Taubenberger19} can utilize the additional cosmological information contained by our full 2D posterior outside the regions probed by our cosmological priors.

For comparison, we also perform cosmological inference using only the 1D posterior of $\Dd$ (dashed contours in right panel of \figref{fig:h0}). This gives $\ho = 75.5_{-7.2}^{+8.3}$ \hunit\ (a 10.3\% measurement) for the uniform $\Omega_{\rm m}$-prior, and $\ho = 74.4_{-6.2}^{+8.1}$ \hunit\ (a 9.6\% measurement) for the Pantheon$+$ $\Omega_{\rm m}$-prior. The $\Dd$-only constraints are lower by $\sim$1.4\% ($0.15\sigma$) than that from the full 2D distance posterior (for the uniform $\Omega_{\rm m}$-prior). This slight difference arises from the projection difference of the 2D posterior along the $\Dd$ direction and along the narrow track allowed by our choice of cosmological priors.

\section{Discussion} \label{sec:discussion}

We now compare our results with previous works (\secref{sec:comparison_with_previous}), discuss the improvement of the constraint in this paper over single-aperture stellar kinematics (\secref{sec:improvement_from_resolved}), and describe the limitations of this work (\secref{sec:limitations}).

\subsection{Comparison with previous time-delay \Ho\ measurements} \label{sec:comparison_with_previous}

Our measured value $\ho = \hval$ \hunit\ is consistent with previous measurements from lensing time delays with different treatments of the MSD. These previous studies can be divided into two approaches: the first breaks the MSD by assuming simple parametric mass profiles such as the power law or composite (i.e., NFW halo and stars with constant mass-to-light ratio), and the second breaks the MSD based solely on stellar kinematics. Our study belongs to the second approach by allowing the freedom in the model to be maximally degenerate with \Ho\ and constraining it solely from the spatially resolved stellar kinematics. However, it is illustrative to compare our result with the first approach to discuss the validity of their mass model assumptions.

Following the first approach, \citet{Suyu13, Suyu14} measured \Ho\ = \suyuhval\ \hunit\ from this same system \lensname\ with simple parametric mass profiles using HST imaging. \citet{Chen19} combined the HST imaging and adaptive-optics-assisted imaging from the Keck Telescope to measure \Ho\ = \chenhval\ \hunit\. Although these studies used single-aperture stellar kinematics, the MSD was already broken by the assumption of parametric mass profiles, and the single-aperture velocity dispersion helped tighten the constraint and made the inferred \Ho\ values from the power-law and composite models more consistent\citep{Suyu14}. Our measured value -- albeit with a larger uncertainty due to the maximal freedom allowed in the mass model -- has a median value very close to these previous measurements. Such a good agreement in the medians suggests that these previous studies' simple parametric mass models are close to the ground truth, and no bias is detected within the precision afforded by the data. Future spatially resolved velocity dispersion measurements for more time-delay lens systems or better quality data for this system (e.g., from the \textit{James Webb} Space Telescope) will allow us to make a more definitive statement on the validity of the parametric mass model assumptions.

Following the second approach, \citet{Birrer16} analyzed this same system \lensname\ using HST imaging and single-aperture velocity dispersion. These authors marginalized the effect of MSD by incorporating a source on the prior but found that the \Ho\ posterior strongly depends on the shape of the anisotropy prior. These authors use two different choices for this prior to find \Ho\ = \birrerhvalthislens\ \hunit\ and \Ho\ = \birrerhvalthislensalt\ \hunit. This large difference illustrates that single aperture velocity dispersion imposes only a weak constraint on the anisotropy profile and, thus, on the MSD. This result highlights the need for spatially resolved velocity dispersion, such as the one presented in this study. Our measured \Ho\ has a precision of 9\% while allowing the data to constrain the MSD effect that is maximally degenerate with \Ho, illustrating the power of spatially resolved kinematics in constraining the anisotropy profile and the MSD, despite the seeing-limited nature of our data. In the future, exquisite data from the \textit{James Webb} Space Telescope (JWST) will provide an even more dramatic improvement \citep[4\% \Ho\ precision forecasted,][]{Yildirim21}.

We also compare our result with the measured values of \Ho\ from the current TDCOSMO sample of seven time-delay lenses. With the power-law mass model assumptions, the combination of seven time-delay lenses gives a 2\% measurement with \Ho\ = \millonhval\ \hunit~ \citep{Wong20, Millon20}. However, relaxing this mass profile assumption and constraining the MSD solely from the single-aperture stellar kinematics of the TDCOSMO sample leads to a 9\% uncertainty on the resultant \Ho\ = \birrerhval\ \hunit. In this study, we achieve the same 9\% precision from a single system, highlighting the superb constraining power of spatially resolved kinematics over single-aperture ones.

It is also worth comparing with the result obtained by \citet{Birrer20} when combining the seven TDCOSMO lenses with information obtained from the external SLACS sample of non-time-delay lenses, \Ho\ = 67.4$_{-3.2}^{+4.1}$ \hunit. Given the uncertainties, our new measurement is not statistically inconsistent with that result, although the difference is clearly important from a cosmological standpoint. With the data in hand, we cannot conclude whether (a) the difference is real and the SLACS sample cannot, therefore, be combined with the TDCOSMO sample, or whether (b) it is due to a statistical fluctuation. This study demonstrates that, as we gather more and better data for spatially resolved kinematics and external samples of non-lenses, we will soon be able to conclude whether the difference is real or not.

In the context of the ``Hubble tension'', our new measurement strengthens the tension by reaffirming the previously obtained time-delay \Ho\ measurements that agreed with other local measurement values, e.g., from SH0ES \citep{Riess22}. Although the 9\% uncertainty in \Ho\ from our measurement alone is not sufficient to resolve the tension, it demonstrates that time-delay cosmography can provide a powerful independent perspective with the help of future data from telescopes such as Keck, \textit{JWST}, and the extremely large telescopes \citep[e.g., see forecasts from][]{Shajib18, Yildirim21, Birrer21c}. We cannot help noticing that the median of our measurement is somewhat higher than the mean of the local values ($\sim$73 \hunit). However, the difference is not significant, given the uncertainties. Therefore our likely explanation is that the difference originates from statistical fluctuation pertaining to this system, as the initial \Ho\ measurements using simple parametric assumption all provided such higher values \citep{Suyu13, Suyu14, Birrer16, Chen19}. We conclude by stressing that some dispersion around the mean is, of course, expected, and indeed \citet{Millon20} shows that the seven TDCOSMO lenses scatter around the mean by an amount consistent with the estimated errors.

\begin{figure*}
	\includegraphics[width=\textwidth]{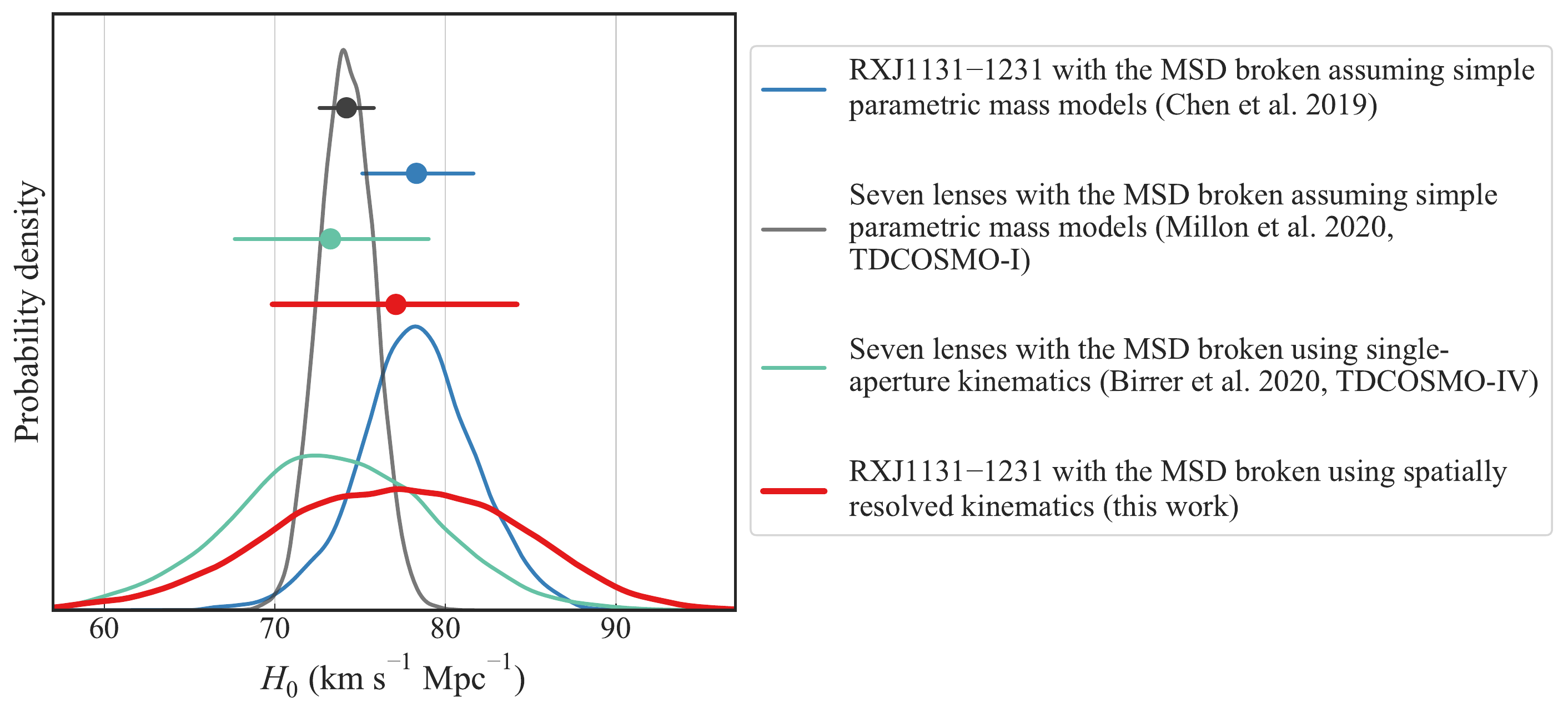}
	\caption{ \label{fig:h0_comparison}
	Comparison of our 9.4\% \Ho\ measurement (red, $\hval$ \hunit) from the single system \lensname\ with previous measurements from \citet[][blue]{Chen19}, \citet[][grey]{Millon20}, and \citet[][emerald]{Birrer20}. The distributions show the \Ho\ posteriors as described in the figure legend, and the points with error bars mark the mean and 64\% credible intervals of the corresponding posterior with matching color. For the same flexible mass models, our analysis on a single system provides a similar precision on \Ho\ with that from seven lenses with only single-aperture stellar kinematics (emerald, \birrerhvalalt \hunit). Moreover, the median value of our measurement falls very close to those from previous analyses on the same system but with simple parametric assumption on the mass model breaking the MSD (blue, \chenhval \hunit, \textit{cf.} also \suyuhval\ \hunit by \citealt{Suyu14}).
	}
\end{figure*}

\subsection{Improvement from the spatial resolution of the stellar kinematics} \label{sec:improvement_from_resolved}

We investigate the improvement in constraints provided by the spatially resolved nature of the stellar kinematics presented in this paper over the unresolved or single-aperture case. \citet{Suyu13} presents a single-aperture measurement of the line-of-sight velocity dispersion $\sigma_{\rm los} = 323 \pm 20$ \kmps\ obtained within a $0\farcs81 \times 0\farcs7$ aperture with a $0\farcs7$ seeing. This measurement was from the Low-Resolution Imaging Spectrometer \citep[LRIS;][]{Oke95} on the Keck Observatory. The probed wavelength range was $\sim$3900--4700 \AA, which probes mostly the redward range of the Ca H\&K lines with a little overlap with the range probed by our data (i.e., 3300--4200 \AA). If we take a luminosity-weighted-sum of the spatially resolved velocity dispersion map within the same $0\farcs81 \times 0\farcs7$ aperture, we get $288 \pm 5$ \kmps, which is $1.7\sigma$ (11\%) lower than the previous single-aperture measurement. Although the $1.7\sigma$ difference is not statistically significant, some parts of it can be due to potential systematics in the kinematic extraction procedure or due to different wavelength ranges probed. It is generally considered that the minimum error, considering systematics, on velocity dispersion measurements is 5\%, even for very high-$S/N$ data.

\begin{figure}
	\includegraphics[width=\columnwidth]{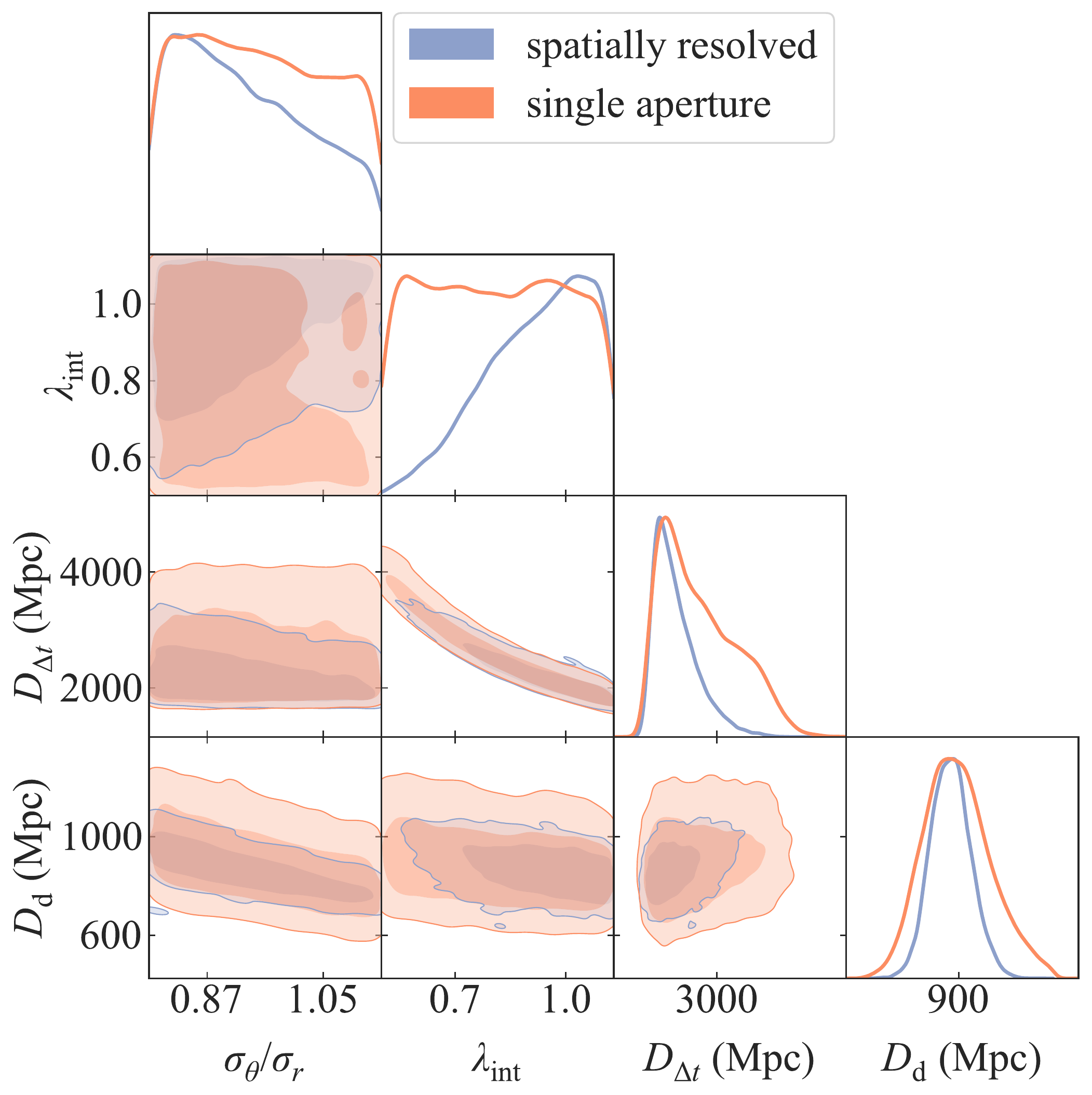}
	\caption{\label{fig:ifu_vs_slit_compare}
	Comparison of the distance constraints between spatially resolved velocity dispersion and single-aperture velocity dispersion. Here, the integrated velocity dispersion is taken as the fiducial value of $287 \pm 18$ \kmps\ to match the mean of our spatially resolved measurement, but the uncertainty of a single-aperture velocity dispersion measurement \citep{Suyu13}. The darker and lighter shaded regions in the 2D plots trace 68\% and 95\% credible regions, respectively. The single-aperture velocity dispersion cannot constrain the anisotropy profile parameter $\sigma_{\theta}/\sigma_{r}$ and the internal MST parameter $\lambda_{\rm int}$, with both limited by the prior. As a result, the $\Ddt$--$\Dd$ posterior is constrained much more weakly.
	}
\end{figure}

However, to illustrate the improvement in precision from the spatially resolved nature of the velocity dispersion presented in this study, we take a fiducial single-aperture measurement value of $288 \pm 18$ \kmps. This mean value is from the luminosity-weighted sum within the single aperture mentioned above, and the 18 \kmps\ uncertainty comes from applying the 6\% uncertainty of the $323 \pm 20$ \kmps\ measurement on the fiducial mean. We take the galaxy's major axis to align with the rectangular aperture's longer side. Rotating the aperture by $90\degr$ only changes the predicted velocity dispersion integrated within the aperture by $\lesssim0.1$\%, which is unsurprising given the mild ellipticity ($q_{\rm l} \sim 0.85$) of the galaxy and the $0\farcs96$ seeing. We compare the key dynamical model parameters between the spatially resolved and single-aperture cases in \figref{fig:ifu_vs_slit_compare}. As expected, the internal MST parameter $\lambda_{\rm int}$ and the anisotropy profile parameter $\sigma_{\theta}/\sigma_{r}$ are almost completely unconstrained in the case of the single-aperture stellar kinematics due to the mass-anisotropy degeneracy \citep{Treu02b, Courteau14}. However, the angular diameter distance $D_{\rm d}$ can be constrained to 15.7\% precision, largely by the anisotropy prior (\textit{cf.} the 9.6\% constraint on $\Dd$ from the spatially resolved data). This single-aperture precision level on $\Dd$ agrees very well with the 17.9\% precision on $\Dd$ ($= 810_{-130}^{+160}$ Mpc) obtained by \citet{Jee19} from the same system \lensname\ based on the previously available single-aperture stellar kinematics mentioned above. The Hubble constant \Ho\ can be inferred to 12.5\% precision with the uniform $\Omega_{\rm m}$ prior from the full 2D posterior of the fiducial single-aperture case. Although the improvement in \Ho\ precision (by $\sim$3\%) from the spatially resolved kinematics does not appear to be dramatic, this is due to the fact that the projection of $\Ddt$--$\Dd$ posterior along the narrow track allowed by our chosen prior happens to give a small difference between the two cases. The improvement could have appeared more drastic if the full 2D posterior had a different orientation from the prior region. In reality, the full cosmological information (illustrated by the area enclosed within the 95\% contour) contained by the single-aperture data is much diluted than that from the spatially resolved data presented in this study (see the $\Ddt$--$\Dd$ contours in \figref{fig:ifu_vs_slit_compare}).

\subsection{Limitations of this study} \label{sec:limitations}

One limitation of our study is the data quality. Although our data are the first of their kind from a cutting-edge ground-based facility such as the Keck Observatory, there are opportunities to obtain better-quality data. The KCWI instrument is seeing-limited. Thus the $S/N$ on the lensing galaxy is degraded by contamination from the nearby quasars, and the spatial resolution of the velocity dispersion map is limited by the seeing. Adaptive-optics-assisted IFU spectroscopy from the ground or observations from space, e.g., with the JWST, can deliver exquisite spatially resolved data for improved \Ho\ precision in the future \citep{Yildirim20, Yildirim21}.

Future data with higher spatial resolution will be particularly powerful in constraining the anisotropy profile better. Our measurement has only weak constraints on the anisotropy profile, which is largely bounded by the adopted uniform prior (see \figref{fig:axisymmetric_jam_corner_powerlaw}). This prior is obtained from a sample of eight local massive ellipticals with one of the highest quality spatially resolved kinematics. However, this is a small sample size. A tighter anisotropy prior from larger samples of massive ellipticals, even better if they are from a redshift range that matches with the one for our system, will be helpful to mitigate further the degeneracy induced by the anisotropy profile, i.e., the mass-anisotropy degeneracy \citep{Treu02b, Courteau14}. 

\section{Conclusion} \label{sec:conclusion}

We measured the spatially resolved stellar velocity dispersion of the lens galaxy in \lensname\ using the KCWI IFU spectrograph on the Keck Observatory. We combined the new spatially resolved stellar kinematics with previously obtained lens models derived from HST imaging data, observed time delays, and estimated line-of-sight lensing effects (i.e., the external convergence) to infer \Ho. Combining the spatially resolved velocity dispersion with lens imaging and time delays simultaneously alleviates the MSD in the measured $\Ddt$ and additionally measures the angular diameter distance $\Dd$.

In order to prevent conscious or unconscious experimenter bias, we blindly performed the dynamical modeling and the cosmographic inference. We unblinded the \Ho\ value after all the co-authors had agreed on the modeling choices after various checks on systematics, and the analysis was frozen. The main conclusions from our study are as follows:

\begin{itemize}
    \item The 2D distance posterior of $\Dd$ and $\Ddt$ gives $\ho = \hval$ \hunit\ for a uniform prior on $\Omega_{\rm m} \sim \mathcal{U}(0.05, 0.5)$, and $\ho = 76.0_{-6.6}^{+7.3}$ \hunit for a Gaussian prior on $\Omega_{\rm m}$ from the Pantheon$+$ analysis \citep{Brout22}.
    \item Our 9.4\% measurement from a single system with spatially resolved kinematics provides a similar precision as, and is in excellent agreement with, the current TDCOSMO sample of seven time-delay lenses based only on single-aperture stellar kinematics \citep[\Ho\ = \birrerhval\ \hunit,][]{Birrer20}. Note that the system \lensname\ analyzed here is part of that sample of seven.
    \item The median value of \Ho\ from our analysis is very close to the previously inferred values assuming simple parametric mass models \citep[e.g., \Ho\ = \chenhval\ \hunit,][]{Chen19}. Thus we do not detect any potential bias in those mass profile assumptions within the precision afforded by our data.
\end{itemize}

In conclusion, our study provides an important validation of previous work by our collaboration on the determination of \Ho\ from time-delay cosmography. This analysis also showcases the power of spatially resolved kinematics in breaking the degeneracies that limit the \Ho\ precision when mass profile assumptions on the galaxy density profile are relaxed. As the first application of such methodology performed on real data, this study stands as an important proof of concept to pioneer future studies on many more time-delay lens systems. A future sample of $\sim$40  lensed quasars can independently provide $\sim$1.2\% precision on \Ho\ that is necessary to resolve or confirm the ``Hubble tension'' at 5$\sigma$ confidence level, thanks to spatially resolved stellar kinematics \citep{Birrer21b}.

%--------------------------------------------------------------------
%\facility{Keck:I (LRIS)}
%\facility{Keck:II (ESI)}

\begin{acknowledgements}
%AJS
We thank Elizabeth Buckley-Geer, Thomas E.~Collett, Philip J.~Marshall, and Chiara Spiniello for useful discussions and comments that improved this study and the manuscript.

Support for this work was provided by NASA through the NASA Hubble Fellowship grant HST-HF2-51492 awarded to AJS by the Space Telescope Science Institute, which is operated by the Association of Universities for Research in Astronomy, Inc., for NASA, under contract NAS5-26555.
%GC-FC
TT and GCFC acknowledge support by NSF through grants NSF-AST-1906976 and NSF-AST-1836016, and from the Moore Foundation through grant 8548.
SHS thanks the Max Planck Society for support through the Max Planck Research Group and the Max Planck Fellowship. SHS is supported in part by the Deutsche Forschungsgemeinschaft (DFG, German Research Foundation) under Germany's Excellence Strategy - EXC-2094 - 390783311. 
% DS and FC
This project has received funding from SNSF and from the European Research
Council (ERC) under the European Union’s Horizon
2020 research and innovation programme (COSMICLENS :
grant agreement No 787886)
VNB gratefully acknowledges assistance from National Science Foundation (NSF) Research at Undergraduate Institutions (RUI) grant AST-1909297. Note that findings and conclusions do not necessarily represent views of the NSF.
\\
This work used computational and storage services associated with the Hoffman2 Shared Cluster provided by UCLA Institute for Digital Research and Education’s Research Technology Group.
\\
The data presented herein were obtained at the W. M. Keck Observatory, which is operated as a scientific partnership among the California Institute of Technology, the University of California and the National Aeronautics and Space Administration. The Observatory was made possible by the generous financial support of the W. M. Keck Foundation. The authors wish to recognize and acknowledge the very significant cultural role and reverence that the summit of Maunakea has always had within the indigenous Hawaiian community.  We are most fortunate to have the opportunity to conduct observations from this mountain.
\\
This research made use of \textsc{jampy} \citep{Cappellari08, Cappellari20}, \textsc{pPXF} \citep{Cappellari17, Cappellari22}, \textsc{pafit} \citep{Krajnovic06}, \textsc{vorbin} \citep{Cappellari03}, \textsc{mgefit} \citep{Cappellari02}	, \textsc{lenstronomy} \citep{Birrer18, Birrer21b}, \textsc{numpy} \citep{Oliphant15}, \textsc{scipy} \citep{Jones01}, \textsc{astropy} \citep{AstropyCollaboration13, AstropyCollaboration18}, \textsc{jupyter} \citep{Kluyver16}, \textsc{matplotlib} \citep{Hunter07}, \textsc{seaborn} \citep{Waskom14}, \textsc{emcee} \citep{Foreman-Mackey13}, and \textsc{getdist} (\url{https://github.com/cmbant/getdist}).
\end{acknowledgements}

\bibliographystyle{aa} % style aa.bst
\bibliography{ajshajib} % your references Yourfile.bib

\end{document}